\documentclass[letterpaper, 10 pt, conference]{ieeeconf}  

\IEEEoverridecommandlockouts                              
                                                          
\usepackage{graphics} 
\usepackage{times} 
\usepackage{amsmath} 
\usepackage{amssymb}  
\usepackage{booktabs}
\usepackage{url}
\usepackage{hyperref}
\usepackage{xcolor}
\usepackage{float}
\usepackage{optidef}
\usepackage{cite}
\usepackage{mathtools}
\usepackage{nomencl}
\usepackage{array}
\usepackage{caption}
\usepackage{subcaption}

\usepackage{amsthm}
\theoremstyle{remark}
\newtheorem{remark}{Remark}
\usepackage{multirow}
\usepackage[ruled,vlined]{algorithm2e}
\usepackage{optidef}
\usepackage{hhline}
\usepackage{array}
\usepackage{amsthm}
\usepackage{multirow}
\usepackage{soul}
\usepackage{nomencl}
\makenomenclature
\title{\LARGE \bf Estimation of Cell-to-Cell Variation and State of Health for Battery Modules with Parallel-Connected Cells}

\author{Qinan Zhou$^{1,*}$, and Jing Sun$^{2}$
\thanks{$^{1}$Qinan Zhou is with the Department of Mechanical Engineering, University of Michigan, Ann Arbor, MI 48103, USA. Email: {\tt\small qinan@umich.edu}}%
\thanks{$^{2}$Jing Sun is with the Department of Naval Architecture and Marine Engineering, University of Michigan, Ann Arbor, MI 48103, USA. Email: {\tt\small jingsun@umich.edu}}%
\thanks{$^{*}$Corresponding Author.}%
}

\begin{document}

\maketitle
\thispagestyle{plain}
\pagestyle{plain}
\begin{abstract}
Estimating cell-to-cell variation (CtCV) and state of health (SoH) for battery modules composed of parallel-connected cells is challenging when only module-level signals are measurable and individual cell behaviors remain unobserved. Although progress has been made in SoH estimation, CtCV estimation remains unresolved in the literature. This paper proposes a unified framework that accurately estimates both CtCV and SoH for modules using only module-level information extracted from incremental capacity analysis (ICA) and differential voltage analysis (DVA). With the proposed framework, CtCV and SoH estimations can be decoupled into two separate tasks, allowing each to be solved with dedicated algorithms without mutual interference and providing greater design flexibility. The framework also exhibits strong versatility in accommodating different CtCV metrics, highlighting its general-purpose nature. Experimental validation on modules with three parallel-connected cells demonstrates that the proposed framework can systematically select optimal module-level features for CtCV and SoH estimations, deliver accurate CtCV and SoH estimates with high confidence and low computational complexity, remain effective across different C-rates, and be suitable for onboard implementation. 
\end{abstract}

\nomenclature[01]{$\mathcal{A}$}{Set of All Features}
\nomenclature[02]{$F, G, H$}{Random Variable}
\nomenclature[03]{$I$}{Mutual Information or Conditional Mutual Information}
\nomenclature[04]{$\tilde{I}$}{Normalized Mutual Information or Normalized Conditional Mutual Information}
\nomenclature[05]{$i, j, l$}{Index}
\nomenclature[06]{$K$}{Kernel Function}
\nomenclature[07]{$N$}{Total Number of Sample Points}
\nomenclature[08]{$Q$}{Charged Capacity}
\nomenclature[09]{$\mathcal{R}$}{Removed Feature Set}
\nomenclature[10]{$\mathcal{S}$}{Ranked Selected Feature Set}
\nomenclature[11]{$\mathcal{U}$}{Set of Features Not Selected Yet}
\nomenclature[12]{$V$}{Voltage}
\nomenclature[13]{$\boldsymbol{w}$}{Weight Vector}
\nomenclature[14]{$X$}{Feature Random Variable}
\nomenclature[15]{$\boldsymbol{x}$}{Input Vector}
\nomenclature[16]{$Y$}{Output Random Variable}
\nomenclature[17]{$\boldsymbol{y}$}{Vector Containing All Outputs in the Dataset}
\nomenclature[18]{$Z$}{Any Random Variables or Combinations of Random Variables}
\nomenclature[19]{$\boldsymbol{\alpha}$}{Vector with $\alpha_i$ where $\alpha_i^{-1}=$ Variance of $i$-th Weight Distribution}
\nomenclature[20]{$\beta$}{Scalar where $\beta^{-1} =$ Variance of Noise Distribution}
\nomenclature[21]{$\boldsymbol{\mu}$}{Mean Vector}
\nomenclature[22]{$\boldsymbol{\Phi}$}{Kernel Matrix}
\nomenclature[23]{$\boldsymbol{\Sigma}$}{Covariance Matrix}
\printnomenclature
\section{INTRODUCTION}
\label{Intro}
Lithium-ion battery cells are connected in series and parallel to form battery modules for practical applications \cite{PlettBook}. Monitoring the degradation of battery cells and modules from onboard measurements is crucial for ensuring accurate range estimation, reliable performance, safety, maintenance scheduling, and warranty management \cite{Noura}. 

At the cell level, a key degradation monitoring task is the state of health (SoH) estimation \cite{Noura}. SoH can be characterized by either capacity fading or resistance growth \cite{Berecibar,CalRegulation}. This paper focuses on capacity fading. Existing cell-level SoH (C-SoH) estimation approaches can be categorized into model-based and data-driven methods \cite{Noura,Berecibar,Yao}. As a promising middle ground between these two categories, incremental capacity analysis (ICA) and differential voltage analysis (DVA) embed degradation physics into data-driven frameworks through physically interpretable incremental capacity (IC) and differential voltage (DV) features, without requiring complex mechanistic models \cite{Weng1,Weng2,Weng3,Wang,Zhou1,Dubarry,Krupp,Stephens}.

\begin{figure*}[ht]
    \centering
    \includegraphics[width=0.99\textwidth,trim=20 5 8 8,clip]{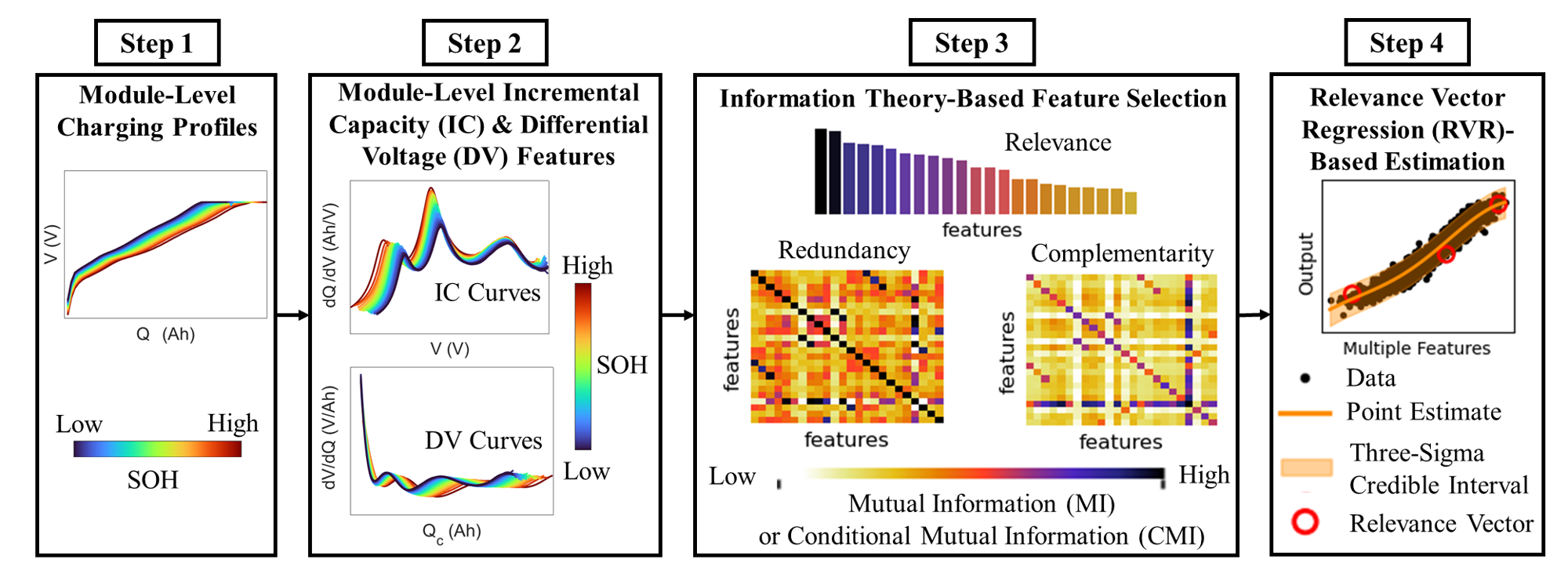}
    \caption{Overview of Proposed Framework for Estimation of Cell-to-Cell Variation and Module-Level State of Health}
    \label{fig:Algorithm}
\end{figure*} 

At the module level, the battery modules composed of parallel-connected cells are the focus of this paper, as their degradation monitoring faces two primary challenges, compared with the cell level: (i) the inevitable cell-to-cell variations (CtCVs) within battery modules \cite{Baumann}, and (ii) the lack of cell-level measurements \cite{PlettBook}. CtCVs can manifest in many forms, including inhomogeneities in cell-level internal impedances, contact resistances, capacities, temperatures, SoH, state of charge (SoC), and other electrochemical or physical properties \cite{Gong, Baumann}. Among these inhomogeneities, this paper focuses specifically on the CtCV in C-SoH values. CtCVs are critical at the module level because they govern internal current sharing, thermal imbalance, and uneven aging among parallel-connected cells, directly impacting the module’s usable capacity, safety margins, and lifetime \cite{Liu2019,Kim2023}. As a result, module-level degradation monitoring must estimate both the module-level SoH (M-SoH) and the CtCV, relying solely on the limited information contained in module-level measurements.

Most existing studies in the literature focus on understanding the root causes \cite{Beck}, effects \cite{Liu2019, Kim2023, Weng4}, and evolution \cite{Gong,Baumann,Lu2024,Wildfeuer,Brand,Fernández,Song,Song2} of CtCVs, whereas very few have addressed the challenge of the CtCV estimation using only module-level measurements. For their effects, CtCVs lead to nonuniform current and temperature distributions among cells within battery modules, thereby distorting the correlation between module-level signals and the true degradation status of modules \cite{Liu2019, Kim2023, Weng4}.

The evolution of CtCVs as battery modules age remains debated. Some studies report divergence driven by nonuniform thermal fields, contact and interconnect resistances, and heterogeneous operating conditions \cite{Gong,Baumann,Lu2024,Wildfeuer}, whereas others suggest convergence due to a self-balancing effect in which healthier cells carry more load, age faster, and gradually drift toward the module average \cite{Brand,Fernández,Song,Song2}. However, the rate of convergence remains poorly characterized in the literature and can be potentially slow. For example, cells with initially identical C-SoH at 100\% can diverge initially due to CtCVs in other cell-level properties and still differ by 5\% in C-SoH when the M-SoH drops to 65\% \cite{Song2}. Hence, module-level degradation monitoring should continue to assume the presence of CtCVs.

For M-SoH estimation, CtCV-induced uneven current distributions distort module-level signals, making conventional cell-level methods ineffective \cite{An}. Early studies attempted to identify CtCV-insensitive, SoH-correlated features to directly apply cell-level estimators at the module level \cite{Weng4,Wang2}. However, such insensitivity is largely phenomenological and may not generalize across battery chemistries or charging conditions \cite{Zhou2}. Consequently, recent studies have shifted toward explicitly addressing CtCV effects through feature selection \cite{Zhou2} or deep learning \cite{Tang,Zhou3,Fan}.

For CtCV estimation using only module-level signals, the literature remains sparse and largely exploratory. Existing studies primarily demonstrate feasibility rather than provide functional estimators. Typically, the existing work simulates artificial modules with and without CtCVs, and then compares their resulting IC/DV curves under fixed M-SoH and single charging conditions \cite{Wong}. While certain module-level IC/DV (M-IC/DV) features appear correlated with CtCVs under these restricted settings, such observations do not constitute robust estimation methods and may not generalize across operating conditions or to real modules. To the best of our knowledge, no existing work has proposed a functioning, experimentally validated CtCV estimation method that applies across different M-SoHs and charging C-rates.

This paper first identifies and adopts CtCV metrics to enable numerical estimation. Then, continuing from the work of \cite{Zhou2}, this paper proposes a unified ICA/DVA-based degradation monitoring framework, illustrated in Fig. \ref{fig:Algorithm}, that estimates both CtCV and M-SoH for battery modules with parallel-connected cells under various charging C-rates, using only module-level measurements.

\begin{figure*}[ht]
    \centering
    \begin{subfigure}[h]{0.3\textwidth}
        \centering
        \includegraphics[width=\textwidth,trim=10 1 40 25,clip]{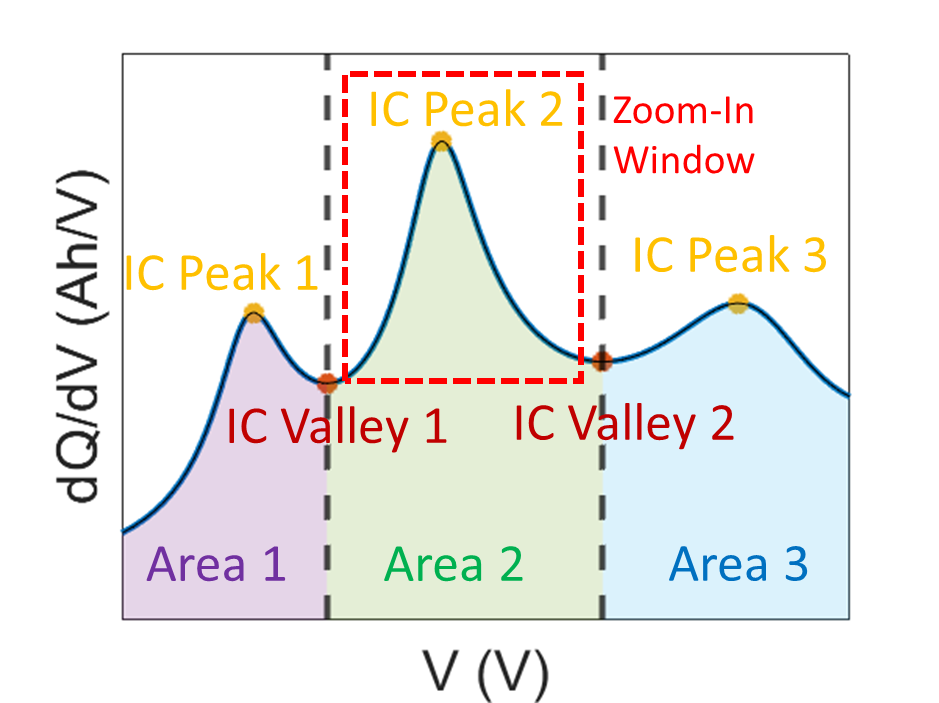}
        \caption{IC Curve}
        \label{fig:IC_Curve_F1}
    \end{subfigure} \hfill
    \begin{subfigure}[h]{0.3\textwidth}
        \centering
        \includegraphics[width=\textwidth,trim=10 1 40 25,clip]{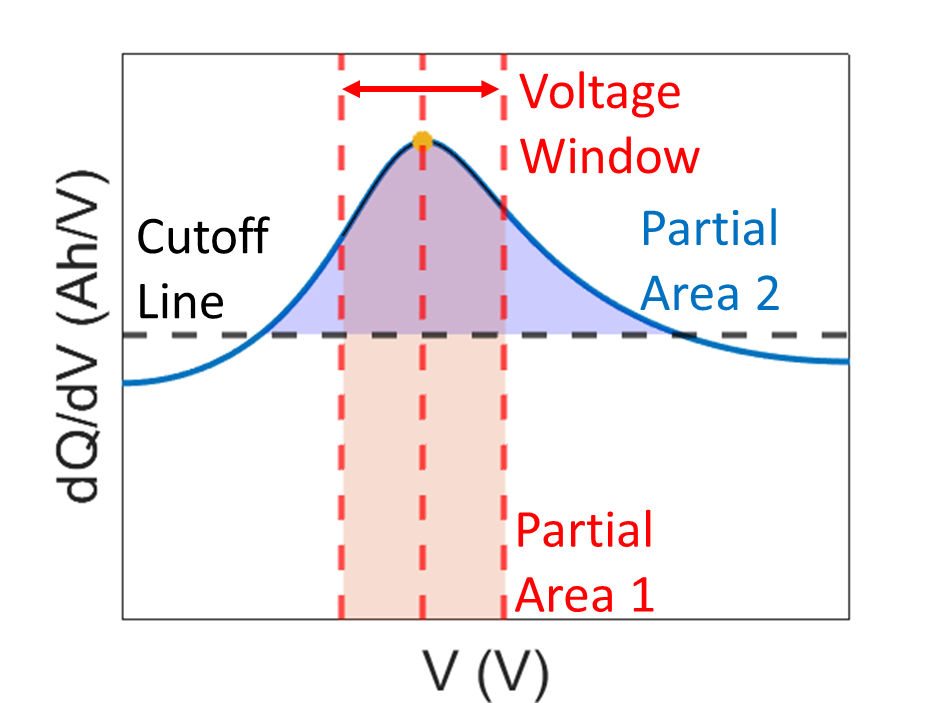}
        \caption{Zoomed-In IC Curve at Example Peak}
        \label{fig:Zoomed_In_IC_Curve_F2}
    \end{subfigure} \hfill
    \begin{subfigure}[h]{0.3\textwidth}
        \centering
        \includegraphics[width=\textwidth,trim=10 1 40 25,clip]{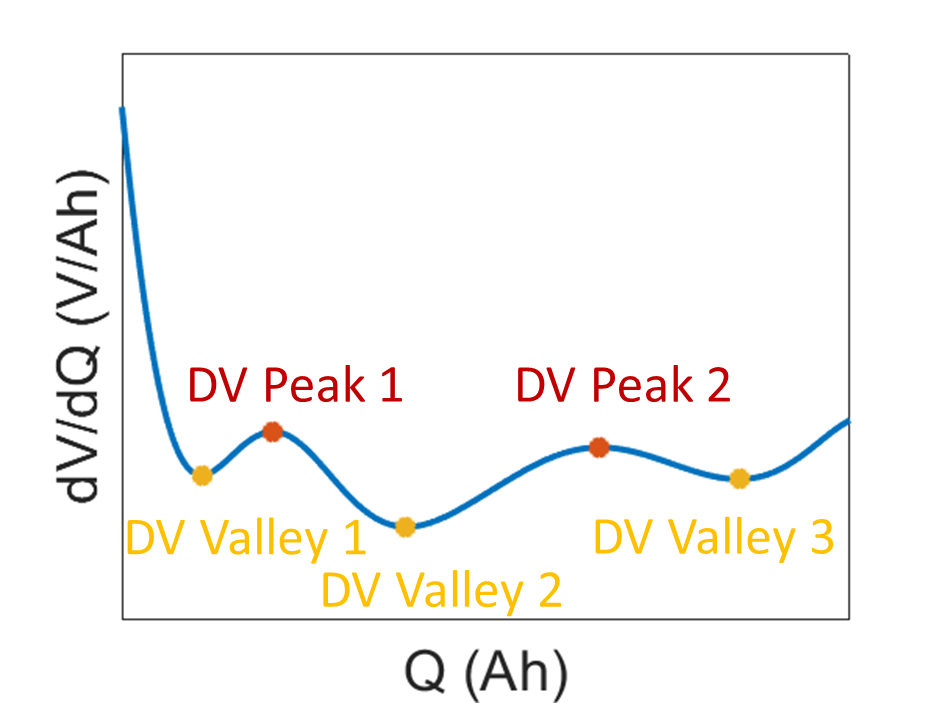}
        \caption{DV Curve}
        \label{fig:DV_Curve_F3}
    \end{subfigure} 
\caption{Definitions of Incremental Capacity (IC) and Differential Voltage (DV) Features}
\label{fig:Feature_Def}
\end{figure*}

For key contributions of this work, on one hand, the proposed framework extends the prior M-SoH estimation approach \cite{Zhou2} to different charging conditions, such as different C-rates. On the other hand, and more importantly, the proposed framework represents the first experimentally validated method in the literature that is capable of performing quantitative CtCV estimation across different M-SoH values and charging C-rates. Specifically, the contributions of the paper are three-fold. 
\begin{itemize}
    \item First, this paper demonstrates that the proposed framework can accurately estimate both CtCV and M-SoH from module-level signals under different charging C-rates, while maintaining low computational complexity.
    \item Second, this paper shows that CtCV and M-SoH can be independently estimated using the proposed framework, providing greater flexibility in algorithm design without mutual interference. 
    \item Third, the performance of the proposed framework has been validated on an experimental dataset of lithium nickel-cobalt-aluminum oxide (NCA) modules consisting of three parallel-connected cells.
\end{itemize} 


To elucidate the proposed framework (Fig. \ref{fig:Algorithm}), the paper proceeds as follows. Section \ref{Concepts} introduces the underpinning definitions and analyzes how CtCV manifests as distortions in module-level signals. Section \ref{Framework} develops the proposed unified framework for CtCV and SoH estimations. Section \ref{Dataset} describes the experimental datasets, while Section \ref{Results} evaluates the performance of the proposed framework. Finally, Section \ref{Conclusion} summarizes the paper.
\section{FOUNDATIONAL CONCEPTS}
\label{Concepts}
This section introduces foundational definitions that underpin the rest of this work. First, it presents CtCV and SoH metrics proposed in the literature and adopts those suitable for quantitative estimation. Second, it explains the formulation of M-IC/DV curves and their associated features. Finally, it characterizes the CtCV-induced distortions in M-IC/DV curves and features, highlighting both the feasibility and challenges of leveraging these distorted M-IC/DV features for CtCV and M-SoH estimations.

\subsection{Quantification of Cell-to-Cell Variation and Module-Level SoH}
\label{Metric}
For SoH metrics, this paper focuses on the capacity fading. Following the standard definition \cite{PlettBook}, this paper defines the C-SoH as: 
\begin{equation}
    \text{C-SoH} = \frac{C_c}{C_{c,\text{fresh}}}, \label{eqn:CSoH_Definition}
\end{equation}
where $C_c$ and $C_{c,\text{fresh}}$ denote the current and fresh cell-level capacities, respectively. Similarly, the M-SoH can be defined as: 
\begin{equation}
    \text{M-SoH} = \frac{C_m}{C_{m,\text{fresh}}}, \label{eqn:MSoH_Definition}
\end{equation}
where $C_m$ and $C_{m,\text{fresh}}$ denote the current and fresh module-level capacities, respectively. Assuming all cells within a module have the same fresh cell-level capacity, M-SoH defined in Equation (\ref{eqn:MSoH_Definition}) can also be expressed as:
\begin{equation}
    \text{M-SoH} = \frac{\sum_{i=1}^{N_p} C_{c,i}}{N_p C_{c,\text{fresh}}} = \text{mean}\left(  \left\{ \text{C-SoH}_i \right\}  \right), \label{eqn:equivalent_MSOH_Def} 
\end{equation}
where $N_p$ is the number of parallel-connected cells, $C_{c,i}$ denotes the cell-level capacity of the $i$-th cell inside the module, and $\left\{ \text{C-SoH}_i \right\} = \left\{ \text{C-SoH}_1, \text{C-SoH}_2, ..., \text{C-SoH}_{N_p} \right\}$. If cells exhibit slight variations in fresh capacity, $C_{c,\text{fresh}}$ can be defined as the maximum fresh capacity among the cells in the module: $C_{c,\text{fresh}} = \text{max}\left(  \left\{ C_{c,i,\text{fresh}} \right\}  \right) $. Accordingly, C-SoH definition given as Equation (\ref{eqn:CSoH_Definition}) is adjusted, while M-SoH definition given as Equation (\ref{eqn:equivalent_MSOH_Def}) remains valid.

While the standard metrics have been used for SoH, there is no universally accepted metric for quantifying CtCVs. A wide range of CtCV metrics have been proposed, such as standard deviation (SD), coefficient of variation, range, skewness, kurtosis, etc. \cite{Wildfeuer, Fernández, Lu2020, Wong}. This paper focuses on the CtCV in C-SoH values and adopts the population SD as the metric, defined as:
\begin{equation}
    \text{SD} = \text{sd} \left( \left\{ \text{C-SoH}_i \right\} \right), 
\end{equation}
due to its simplicity, interpretability, popularity, and generalizability to modules with an arbitrary number of parallel-connected cells. Section \ref{DataOverview} will provide specific examples to build intuitions about the scale of SD as a CtCV metric and how SD relates to the underlying C-SoH values within a battery module. 

\subsection{Definitions of Module-Level IC/DV Curves and Features}
\label{Features}
This subsection deals with Step 2 of Fig. \ref{fig:Algorithm} where all the available features in M-IC/DV curves are extracted. Given measured module-level $Q$-$V$ (charged capacity v.s. voltage) profiles under constant-current (CC) charging, IC and DV are defined as $\text{IC} = dQ/dV$ and $\text{DV} = dV/dQ$, respectively. In this study, support vector regression is employed to obtain M-IC/DV curves through analytical differentiation to mitigate noise amplification associated with numerical differentiation \cite{Weng1, Weng2, Weng3}.

The following features, as defined graphically in Fig. \ref{fig:Feature_Def}, can be extracted from M-IC/DV curves. For IC features, as shown in Fig. \ref{fig:IC_Curve_F1}, the locations and heights of IC peaks and valleys are defined as their $x$- and $y$-coordinates, respectively. IC peak areas 1, 2, and 3 are the areas (i) from the minimum voltage to the first IC valley location, (ii) from the first to the second IC valley locations, and (iii) from the second IC valley location to the maximum voltage, respectively \cite{Anseán}. IC peak partial areas can be defined as the area either above a user-defined horizontal cutoff line or within a user-defined symmetric voltage window centered on a particular IC peak \cite{ZhouR}, as shown in Fig. \ref{fig:Zoomed_In_IC_Curve_F2}. Similarly, DV features, as defined in Fig. \ref{fig:DV_Curve_F3}, include the locations ($x$-coordinates) and heights ($y$-coordinates) of DV peaks and valleys. For charging condition features, temperatures and C-rates also influence M-IC/DV curves and therefore serve as essential features for module-level degradation monitoring \cite{Zhou1}. Note that a typical cylindrical cell involves core, surface, and ambient temperatures. Because electrochemical reactions inside the battery govern the shape of IC/DV curves, the core temperature most directly reflects the physics. However, it is rarely measured in onboard applications. Consequently, the surface and ambient temperatures are commonly used as practical proxies. Thus, what temperature is used depends on the desired accuracy in capturing the underlying physics and the availability of measurements.

\begin{table}[H]
\centering
\caption{Summary of Features from ICA, DVA, and Charging Conditions}
\label{table:Features}
\begin{tabular}{|c|c|c|c|}
\hline
\begin{tabular}[c]{@{}c@{}} \textbf{Feature} \\ \textbf{Category} \end{tabular} & \textbf{Feature Name} & \textbf{Acronym} & \textbf{Reference} \\ \hline
\multirow{10}{*}{ \begin{tabular}[c]{@{}c@{}} ICA \\ / \\ DVA \end{tabular} } & IC Peak Height {[}Ah/V{]} & IC PH & \multirow{8}{*}{ \begin{tabular}[c]{@{}c@{}} \cite{Dubarry} \\ \cite{Krupp} \\ \cite{Bloom} \end{tabular} } \\ \cline{2-3}
 & IC Peak Location {[}V{]} & IC PL &  \\ \cline{2-3}
 & DV Valley Height {[}V/Ah{]} & DV VH &  \\ \cline{2-3}
 & DV Valley Location {[}Ah{]} & DV VL &  \\ \cline{2-3}
 & IC Valley Height {[}Ah/V{]} & IC VH &  \\ \cline{2-3}
 & IC Valley Location {[}V{]} & IC VL &  \\ \cline{2-3}
 & DV Peak Height {[}V/Ah{]} & DV PH &  \\ \cline{2-3}
 & DV Peak Location {[}Ah{]} & DV PL &  \\ \cline{2-4} 
 & IC Peak Area {[}Ah{]} & IC AR & \begin{tabular}[c]{@{}c@{}} \cite{Anseán} \end{tabular} \\ \cline{2-4} 
 & IC Peak Partial Area {[}Ah{]} & IC PA & \begin{tabular}[c]{@{}c@{}} \cite{ZhouR} \end{tabular} \\ \hline
\multirow{2}{*}{\begin{tabular}[c]{@{}c@{}} Charging \\ Conditions \end{tabular}} & C-Rate {[}C{]} & - & \multirow{2}{*}{ \begin{tabular}[c]{@{}c@{}} \cite{Zhou1} \end{tabular} } \\ \cline{2-3} 
 & \begin{tabular}[c]{@{}c@{}} (Core, Surface, Ambient) \\ Temperature {[}°C{]} \end{tabular} & - &  \\ \hline
\end{tabular}
\end{table}

Table \ref{table:Features} summarizes all the available features from ICA, DVA, and charging conditions. Their physical interpretations and connections to battery degradation are detailed in the references listed in Table \ref{table:Features}. Note that a key advantage of ICA/DVA over other data-driven methods is that the use of IC/DV features embeds degradation-related physical insights into the monitoring framework without requiring a complex mechanistic model, thereby improving the interpretability.

It is important to recognize that the presence and prominence of these features can vary with battery chemistry and charging conditions. Consequently, some features may disappear, while new features may emerge, leading to a feature set that may differ slightly from the standard case described in this section.

\subsection{Distortions of Module-Level IC/DV Curves due to Cell-to-Cell Variations}
\label{Distortion}
\begin{figure}[H]
    \centering
    \includegraphics[width=0.49\textwidth,trim=45 130 45 20,clip]{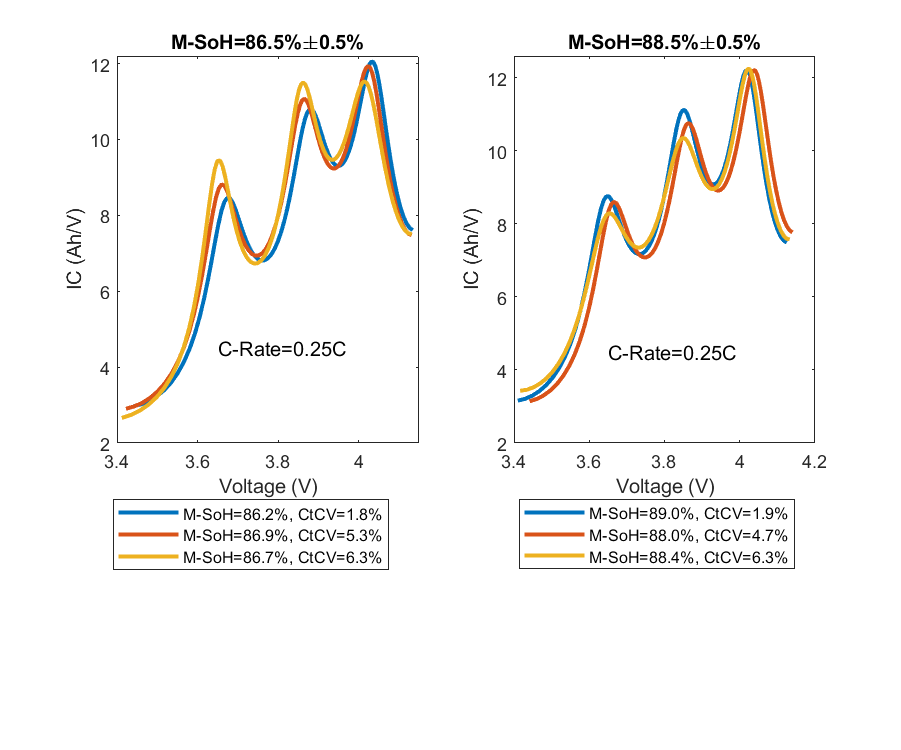}
    \caption{IC Curve Distortions Caused by CtCVs. Note that, under the same C-rate and similar M-SoH, no discernible trend between CtCVs and the resulting IC curve distortions can be observed from the plots.}
    \label{fig:Example_ICcurves_underCtCV}
\end{figure}

CtCVs lead to uneven current distribution across parallel-connected cells within modules, thereby distorting the M-IC/DV curves and their associated features \cite{Weng4}. To demonstrate such distortions, NCA modules composed of three parallel-connected cells are examined. Fig. \ref{fig:Example_ICcurves_underCtCV} presents experimental M-IC/DV curves for NCA modules with different levels of CtCVs at similar M-SoH values of approximately 86.5\% and 88.5\%.

To illustrate how CtCVs distort M-IC/DV features, IC peak height is used as an example. Based on Fig. \ref{fig:Example_ICcurves_underCtCV}, in contrast to the cell level where IC peak heights exhibit a monotonic relationship with C-SoH \cite{Weng1, Weng2, Weng3, Zhou1}, IC peak heights vary significantly for modules with a similar M-SoH due to the presence of CtCVs. Similar phenomena can also be observed for other M-IC/DV features. Thus, these CtCV-induced distortions complicate M-SoH estimation, but can be informative for CtCV estimation.

However, the relationships between CtCVs and distortions in M-IC/DV features become highly nonmonotonic and not easily discerned across different M-SoH values and charging conditions. Using IC peak height as an example, Fig. \ref{fig:Example_ICcurves_underCtCV} shows that the left and middle IC peaks can reverse their trends with respect to CtCVs when M-SoH varies, despite identical charging C-rates. Similar trend reversals can also be observed when the charging C-rate varies. These complex and condition-dependent behaviors highlight the limitations of existing approaches discussed in Section \ref{Intro}: correlations between distorted M-IC/DV features and CtCVs identified under a single M-SoH and charging condition are largely phenomenological and do not generalize reliably to different M-SoH values and charging conditions in experimental data.

As a result, these observations motivate several requirements for a framework that estimates CtCV and M-SoH: (i) it should employ a systematic feature-selection strategy rather than depend on situation-specific phenomenological observations; (ii) it should capitalize on distorted M-IC/DV features for CtCV estimation, but mitigate their adverse effect on M-SoH estimation; and (iii) it should be validated across different charging conditions using experimental data. The proposed framework in Fig. \ref{fig:Algorithm} satisfies all these requirements.
\section{THE PROPOSED FRAMEWORK FOR ESTIMATION OF CELL-TO-CELL VARIATION AND MODULE-LEVEL STATE OF HEALTH}
\label{Framework}
This section addresses Steps 3 and 4 of Fig. \ref{fig:Algorithm}, in which the proposed method for CtCV and M-SoH estimations is developed. 

First, to overcome the limitations of situation-specific, phenomenology-based feature selection (as discussed in Sections \ref{Intro} and \ref{Distortion}), Step 3 of Fig. \ref{fig:Algorithm} employs a generalizable information theory-based feature selection algorithm to identify optimal feature sets for both CtCV and M-SoH estimations. Compared to other feature selection algorithms, this approach is adopted due to its ability to capture nonlinear feature-target relations, well-established selection criteria, computational efficiency, and independence from subsequent estimation models \cite{Li2}. 

Second, with the selected features, Step 4 of Fig. \ref{fig:Algorithm} applies relevance vector regression (RVR), a sparse Bayesian learning paradigm \cite{Tipping01}, to construct probabilistic estimators that provide both point estimates and three-sigma credible intervals for both M-SoH and CtCV. RVR is chosen for its capability to update models based on future data through sequential Bayesian learning, mitigate model uncertainties to enhance robustness, and automatically favor simple but sufficiently accurate models \cite{Tipping01,MacKay}.

\subsection{Key Information-Theoretic Concepts for Feature Selection}
\label{InfoDefn} 

\begin{algorithm*}[h]
\textbf{Output}: the set of selected features $\mathcal{S}$, the set of removed features $\mathcal{R}$ \

\textbf{Input}: set $\mathcal{A}$ containing all the features, threshold $\tilde{I}_{\text{th}}$ for removing completely redundant features\

 \textbf{Initialization}: $\mathcal{S} \leftarrow \emptyset$ or \{pre-selected features\}, $\mathcal{R} \leftarrow \emptyset$, $\mathcal{U}  \leftarrow \mathcal{A} \backslash \mathcal{S} $\

 \

 Find completely redundant features to any pre-selected features: $\mathcal{D} \leftarrow \left\{ X \in \mathcal{U}: \tilde{I}\left( X; X_j \right) \geq \tilde{I}_{\text{th}}, X_j \in \mathcal{S} \right\}$\
 
 $\mathcal{U} \leftarrow \mathcal{U} \backslash \mathcal{D}$, $\mathcal{R}\leftarrow \mathcal{R} \cup \mathcal{D}$ \
 
 \While{$|\mathcal{U}| > 0$}{

 \If{$ |\mathcal{S}| = 0$}{
 $ X^* \leftarrow \operatorname*{argmax}_{X \in \mathcal{U} } \tilde{I}\left(X ; Y \right)$   \
 }
 \Else{
 $ X^* \leftarrow \operatorname*{argmax}_{X \in \mathcal{U} } \left( \tilde{I}\left( X; Y \right) - \frac{1}{|\mathcal{S}|} \sum_{X_j \in \mathcal{S}} \tilde{I} \left(X ; X_j \right) + \frac{1}{|\mathcal{S}|} \sum_{X_j \in \mathcal{S}} \tilde{I} \left(X; X_j | Y \right) \right)$ \
 }
 $\mathcal{S} \leftarrow \mathcal{S} \cup \{ X^* \}$, $\mathcal{U} \leftarrow \mathcal{U} \backslash \{ X^* \}$ \

 Find completely redundant features to $X^*$: $\mathcal{D} \leftarrow \left\{ X \in \mathcal{U} : \tilde{I} \left(X^*; X \right) \geq \tilde{I}_{\text{th}} \right\}$\

 $\mathcal{U}\leftarrow \mathcal{U} \backslash \mathcal{D}$, $\mathcal{R}\leftarrow \mathcal{R} \cup \mathcal{D}$ \
 
 }
\caption{Proposed Information Theory-Based Feature Selection Algorithm}
\label{alg:infoFS}
\end{algorithm*}

Mutual information (MI) and conditional mutual information (CMI) form the cornerstone of the information theory-based feature selection method in Step 3 of Fig. \ref{fig:Algorithm} \cite{Li2}. In general, both quantities are defined for discrete, continuous, or mixed random variables (RVs) \cite{Gao,Mesner}. For three discrete RVs $F$, $G$, and $H$, the MI between $F$ and $G$ is defined as:
\begin{equation}
    I\left(F; G\right) = \sum_{f \in F} \sum_{g \in G} p\left(f,g\right) \log{ \frac{p\left(f,g\right)}{p\left(f\right)p\left(g\right)} }, \label{eqn:MI_def}
\end{equation}
and the CMI between $F$ and $G$ given $H$ are defined as: 
\begin{align}
I\left(F; G | H\right)  
&= \sum_{h \in H} \biggl\{ p\left(h\right) \cdot \notag \\& \sum_{f \in F} \sum_{g \in G} p\left(f,g|h\right) \log{ \frac{p\left(f,g|h\right)}{p\left(f|h\right)p\left(g|h\right)} }  \biggl\}, \label{eqn:CMI_def}
\end{align}
where $p(\cdot)$ is the probability distribution function and $\log(\cdot)$ is the natural logarithm \cite{InfoBook}. If $F$, $G$, and $H$ are continuous RVs, summations are replaced by integrations \cite{Kraskov}. Intuitively, the MI quantifies the uncertainty reduction of an RV if another RV is known, while the CMI measures the uncertainty reduction of an RV if another RV is known given the third RV \cite{InfoBook}. 

Because RVs may exhibit different intrinsic uncertainties, raw MI and CMI values can be difficult to interpret and compare \cite{Estevez}. To address this issue, following \cite{Kvålseth} and \cite{Zhou2}, both quantities are normalized using:  
\begin{eqnarray}
\tilde{I}\left(F;G\right) &=& \frac{I\left(F;G\right)}{\min \left(I\left(F;F\right),I\left(G;G\right)\right)}, \label{eqn:MI_norm}\\ 
\tilde{I}\left(F;G | H\right) &=& \frac{I\left(F;G | H \right)}{\min \left(I\left(F;F\right),I\left(G;G\right)\right)}. \label{eqn:CMI_norm}
\end{eqnarray}

Since the probability distributions in Definitions (\ref{eqn:MI_def}) and (\ref{eqn:CMI_def}) are unknown in practice, MI and CMI must be estimated directly from data. This work adopts the estimator developed by \cite{Mesner} to compute CMI estimates $\hat{I}\left(F;G|H\right)$. Given three RVs $F$, $G$, and $H$, this estimator \cite{Mesner} requires to standardize samples from each RV first using: 
\begin{equation}
    z_i^{(s)} = \frac{z_i - \text{mean}\left( \left\{ z_i \right\}_{i=1}^N \right)}{\text{sd}\left( \left\{ z_i \right\}_{i=1}^N \right)}, \label{eqn:standardize}
\end{equation}
where $Z$ represents an abitrary RV \cite{Runge}. To obtain MI estimates $\hat{I}(F;G)$ that are consistent with the CMI estimates, following \cite{Zhou2}, the same estimator is used by setting $H$ as an independent white Gaussian noise with zero mean and unit variance, yielding: 
\begin{equation}
    \hat{I}\left(F;G\right) = \hat{I}\left(F;G| \text{White Gaussian Noise} \right), 
\end{equation}
based on Definitions (\ref{eqn:MI_def}) and (\ref{eqn:CMI_def}). Additional details on the estimator are provided in \cite{Mesner, Zhou2}. 

\subsection{Information Theory-Based Feature Selection}
\label{FS_Algm}
This subsection discusses the algorithm used in Step 3 of Fig. \ref{fig:Algorithm}. An information theory-based feature selection algorithm typically adopts a greedy, computationally efficient forward sequential search strategy, since identifying a globally optimal set of features is an NP-hard problem \cite{Li2,Ferri}. 

Algorithm \ref{alg:infoFS} presents the pseudo-code for the proposed feature selection algorithm. The notation used in Algorithm \ref{alg:infoFS} is defined as follows: $\mathcal{A}$ -- the set of all features, $\mathcal{S}$ -- the set of selected features, $\mathcal{U}$ -- the set of features not selected yet, $\mathcal{R}$ -- the set of removed features. $\mathcal{S}$ and $\mathcal{R}$ are the outputs of Algorithm \ref{alg:infoFS}. 

At a high level, this algorithm constructs $\mathcal{S}$ by adding one feature at a time based on a specified selection criterion (to be introduced later). The order in which features are added to $\mathcal{S}$ naturally yields a ranking of feature importance for CtCV and M-SoH estimations. This ranking allows practitioners to select subsets that achieve desired trade-offs between estimation accuracy and model complexity.

Specifically, Algorithm \ref{alg:infoFS} begins with an empty set $\mathcal{S} = \emptyset$. Then, at the $l$-th iteration, given $\mathcal{S}_{l-1}$ and $\mathcal{U}_{l-1}$ from the previous iteration, the algorithm solves:
\begin{equation}
    X^*_l = \operatorname*{argmax}_{X \in \mathcal{U}_{l-1} } J\left(X \right), \label{eqn:FS_Optimization}
\end{equation}
where the feature selection criterion $J\left(X \right)$ is modified from the joint mutual information criterion \cite{Meyer} and defined as: 
\begin{multline}
    J\left(X \right) = \tilde{I}\left( X; Y \right) - \frac{1}{|\mathcal{S}_{l-1}|} \sum_{X_j \in \mathcal{S}_{l-1}} \tilde{I} \left(X ; X_j \right) \\ + \frac{1}{|\mathcal{S}_{l-1}|} \sum_{X_j \in \mathcal{S}_{l-1}} \tilde{I} \left(X; X_j | Y \right),
    \label{eqn:FS_obj}
\end{multline}
where $|\cdot|$ is the cardinality of a set, $X \in \mathcal{U}_{l-1}$ is a candidate feature, $X_j \in \mathcal{S}_{l-1}$ are previously selected features ($j$ is an index), and $Y$ is the output (i.e., CtCV or SoH). $J(X)$ includes three terms: relevance $\tilde{I}\left( X; Y \right)$, the average redundancy $1/|\mathcal{S}_{l-1}| \cdot \sum_{X_j \in \mathcal{S}_{l-1}} \tilde{I} \left(X ; X_j \right)$, and the average complementarity $1/|\mathcal{S}_{l-1}| \cdot \sum_{X_j \in \mathcal{S}_{l-1}} \tilde{I} \left(X; X_j | Y \right)$. These terms are explained as follows: 
\begin{itemize}
    \item \textbf{Relevance}: The relevance of $X$ to $Y$ is defined as $\tilde{I}\left( X; Y \right)$. A high relevance indicates that $X$ carries substantial information about $Y$. For CtCV and M-SoH estimations, high-relevance features are strongly correlated with either CtCV or M-SoH, but are insensitive to the other quantity.
    \item \textbf{Redundancy}: The redundancy between $X$ and $X_j$ is given by $\tilde{I} \left(X; X_j \right)$. Large redundancy indicates that $X$ does not contribute much new information to $X_j$ and therefore should be avoided. The total redundancy between $X$ and all previously selected features is defined as $ \sum_{X_j \in \mathcal{S}_{l-1}} \tilde{I} \left(X ; X_j \right)$. The averaging factor $1/|\mathcal{S}_{l-1}|$ brings down the scale as the cardinality of $\mathcal{S}_{l-1}$ grows very fast through iterations \cite{Meyer}. The proposed algorithm minimizes redundancy. 
    \item \textbf{Complementarity}: The complementarity quantifies the synergistic effects between two features to the target \cite{Vergara}. Certain features may exhibit low relevance to $Y$, but become highly informative when considered alongside other features. A representative example is C-rate: by itself, it provides little MI to CtCV or M-SoH, as neither quantity can be inferred solely from the C-rate. However, once other IC/DV features are selected, it becomes critical because it directly influences the values of these features \cite{Zhou1}. Likewise, the complementarity between $X$ and $X_j$ is $\tilde{I} \left(X ; X_j | Y\right)$, the total complementarity is $ \sum_{X_j \in \mathcal{S}_{l-1}} \tilde{I} \left(X ; X_j |Y\right)$, and $1/|\mathcal{S}_{l-1}|$ provides the scaling effect \cite{Meyer}. Maximizing complementarity promotes features that enhance the collective information content among selected features to $Y$. 
\end{itemize}
Optimization (\ref{eqn:FS_Optimization}) is performed until $|\mathcal{U}| =0$. 

As a result, at each iteration, the proposed algorithm identifies a feature that provides the best balance among high relevance, low redundancy, and strong complementarity, not merely based on its correlation with the output. The optimal features can change when situations change, but the proposed framework stays the same. Thus, the proposed framework is generalizable. 

\begin{remark}
    (On Non-Empty Initialization) $\mathcal{S}$ can be initialized either as an empty set to let data speak for themselves or as a set containing pre-selected features informed by physical knowledge.
\end{remark}
\begin{remark}
    (On Removing Features) In CtCV and M-SoH estimation, $\tilde{I} \left(X ; X_j | Y\right)$ can be large for both complementary and purely redundant features, potentially causing undesired selections. To mitigate this, Algorithm \ref{alg:infoFS} includes a removal step: once $X^*$ is selected, any candidate feature $X\in \mathcal{U}$ whose redundancy with $X^*$ exceeds a user-defined threshold $\tilde{I}_{\text{th}}$ is removed from future consideration. Specifically, remove the set $\mathcal{D} = \left\{ X \in \mathcal{U} : \tilde{I} \left(X^*; X \right) \geq \tilde{I}_{\text{th}} \right\}$ from $\mathcal{U}$.
\end{remark}

\subsection{Relevance Vector Regression (RVR)} 
\label{RVR}
This subsection discusses the model used in Step 4 of Fig. \ref{fig:Algorithm}. This work employs the original RVR algorithm \cite{Tipping02}, implemented using the sklearn-RVM package \cite{Tipping03}. For completeness, a brief summary of RVR is provided.

Consider a dataset of independent and identically distributed (I.I.D.) sample $\left\{ \left( \boldsymbol{x}_i, y_i \right) \right\}_{i=1}^N$ where $\boldsymbol{x} \in \mathbb{R}^{N_f}$ are feature vectors, $N_f$ is the number of features, $y \in \mathbb{R}$ is the output (i.e., CtCV or M-SoH), $i$ is an index, and $N$ is the number of sample points. Then, the relationship between an input $\boldsymbol{x}$ and its output $y$ is modeled as:
\begin{equation}
    y = \left( w_0 + \sum_{i=1}^{N} w_i K\left( \boldsymbol{x},\boldsymbol{x}_i \right) \right) + \eta, \label{eqn:RVRmodel}
\end{equation}
where $w_i$ and $w_0$ are weights and offset to be learned, $\eta$ is a white Gaussian noise with an unknown variance $\beta^{-1}$ to be learned, and $K\left(\cdot, \cdot \right)$ is a user-defined kernel \cite{Tipping02}. The radial basis function kernel is adopted here. 

As a Bayesian learning paradigm, RVR first adopts the following hierarchical priors that guarantee model sparsity after training \cite{Tipping02,Robert,Steinke}. 
\begin{eqnarray}
    p\left( \boldsymbol{w}|\boldsymbol{\alpha} \right) &=& \prod_{i=0}^N \mathcal{N}\left( w_i | 0, \alpha_i^{-1} \right), \label{eqn:WeightPrior} \\ 
    p\left(\boldsymbol{\alpha} \right) &=& \prod_{i=0}^N \Gamma\left( \alpha_i | a, b \right), \label{eqn:VariancePrior} \\
    p\left( \beta \right) &=& \Gamma\left(\beta | c,d\right), \label{eqn:NoisePrior}
\end{eqnarray}
where $p(\cdot)$ is the probability distribution function, $\boldsymbol{w} = \begin{bmatrix} w_0 & ... & w_N \end{bmatrix}^T \in \mathbb{R}^{N+1}$ and $\boldsymbol{\alpha} = \begin{bmatrix} \alpha_0 & ... & \alpha_N \end{bmatrix}^T \in \mathbb{R}^{N+1}$ are unknowns to be learned, $\mathcal{N}(\cdot|\cdot, \cdot)$ and $\Gamma(\cdot|\cdot, \cdot)$ are the Gaussian and Gamma distributions respectively, and  $a\rightarrow 0$, $b\rightarrow 0$, $c\rightarrow 0$, $d\rightarrow 0$ \cite{Tipping02}. 

Then, based on the assumptions of I.I.D. sample points and white Gaussian noises in the process, the likelihood for RVR is:  
\begin{equation}
    p\left( \boldsymbol{y} | \boldsymbol{w} ,\beta \right) = \mathcal{N}\left( \boldsymbol{y} | \boldsymbol{\Phi} \boldsymbol{w}, \beta^{-1} \boldsymbol{I}\right),
\end{equation}
where $\boldsymbol{\Phi} = \begin{bmatrix} \boldsymbol{\phi} \left(\boldsymbol{x}_1 \right) & ... & \boldsymbol{\phi} \left(\boldsymbol{x}_N \right) \end{bmatrix}^T \in \mathbb{R}^{N \times \left(N+1\right)}$, $\boldsymbol{\phi} \left(\boldsymbol{x}_i\right) = \begin{bmatrix} 1 & K\left(\boldsymbol{x}_i,\boldsymbol{x}_1\right) & ... & K \left(\boldsymbol{x}_i,\boldsymbol{x}_N\right)\end{bmatrix}^T$, $\boldsymbol{y} = \begin{bmatrix} y_1 & ... & y_N \end{bmatrix}^T$, and $\boldsymbol{I}$ is the identity matrix \cite{Tipping02}.

With the priors and likelihood specified, Bayesian learning proceeds by computing and updating the posterior \cite{Robert}. For the RVR, the posterior is: 
\begin{equation}
    p\left( \boldsymbol{w}, \boldsymbol{\alpha}, \beta | \boldsymbol{y} \right) = p\left( \boldsymbol{w} | \boldsymbol{y},\boldsymbol{\alpha},\beta \right) p \left( \boldsymbol{\alpha},\beta | \boldsymbol{y} \right).
\end{equation}
The first term $p\left( \boldsymbol{w} | \boldsymbol{y},\boldsymbol{\alpha},\beta \right)$ can be found analytically as:
\begin{eqnarray}
    p\left( \boldsymbol{w} | \boldsymbol{y},\boldsymbol{\alpha},\beta \right) &=& \mathcal{N} \left( \boldsymbol{w} | \boldsymbol{\mu}, \boldsymbol{\Sigma} \right), \\
    \boldsymbol{\Sigma} &=& \left( \beta \boldsymbol{\Phi}^T \boldsymbol{\Phi} + \boldsymbol{A} \right)^{-1}, \label{eqn:SigmaMat} \\
    \boldsymbol{\mu} &=& \beta \boldsymbol{\Sigma} \boldsymbol{\Phi}^T \boldsymbol{y}, \label{eqn:muVec}
\end{eqnarray} 
where the diagonal matrix $\boldsymbol{A} = \text{diag}\left( \boldsymbol{\alpha} \right)$. The second term $p \left( \boldsymbol{\alpha},\beta | \boldsymbol{y} \right)$ cannot be analytically evaluated, but can be obtained numerically by iteratively solving the following Type-II maximum likelihood problem \cite{Tipping02}:  
\begin{eqnarray}
\boldsymbol{\alpha}_{\text{MP}}, \beta_{\text{MP}} &=& \operatorname*{argmax}_{\boldsymbol{\alpha},\beta} p\left(\boldsymbol{y}|\boldsymbol{\alpha},\beta \right), \label{eqn:maximum_likelihood_opt}\\
p\left(\boldsymbol{y}|\boldsymbol{\alpha},\beta\right) &=& \mathcal{N}\left( \boldsymbol{y} | \boldsymbol{0}, \beta^{-1} \boldsymbol{I} + \boldsymbol{\Phi}\boldsymbol{A}^{-1} \boldsymbol{\Phi}^T \right),
\end{eqnarray}
where $\boldsymbol{0}$ is a vector with all components equal to 0.

As iteration progresses for solving Optimization (\ref{eqn:maximum_likelihood_opt}), most of the $\alpha_i$ approach $\infty$, causing their associated weights $w_i$ to collapse to zero-mean, zero-variance distributions. When this happens, the corresponding entries in $\boldsymbol{\Sigma}$, $\boldsymbol{\mu}$, $\boldsymbol{\Phi}$, and $\boldsymbol{\alpha}$ are deleted/pruned from subsequent iterations \cite{Tipping02}. The $\boldsymbol{x_i}$ related to the remaining $w_i$ with nonzero posterior variance constitute the relevance vectors \cite{Neal}. 

When predicting outputs from the trained RVR \cite{Tipping02}, given a new input $\boldsymbol{x}$, the distribution of estimated output $y$ is: 
\begin{eqnarray}
    p\left( y | \boldsymbol{y},\boldsymbol{\alpha}_{\text{MP}},\beta_{\text{MP}} \right) &=& \mathcal{N} \left( y | t,\sigma^2 \right), \\
    t &=& \boldsymbol{\mu}^T \boldsymbol{\tilde{\phi}} \left(\boldsymbol{x} \right), \label{eqn:mean_estimate}\\
    \sigma^2 &=& \beta^{-1}_{\text{MP}} + \boldsymbol{\tilde{\phi}}\left(\boldsymbol{x} \right)^T \boldsymbol{\Sigma} \boldsymbol{\tilde{\phi}} \left(\boldsymbol{x} \right), \label{eqn:var_estimate}
\end{eqnarray}
where $\boldsymbol{\tilde{\phi}} \left(\boldsymbol{x} \right) = \begin{bmatrix} 1 & K\left( \boldsymbol{x}, \textbf{rv}_1 \right) & ... & K\left( \boldsymbol{x}, \textbf{rv}_{N_{\text{rv}}}\right) \end{bmatrix}^T$ if the offset is used and $\boldsymbol{\tilde{\phi}}\left(\boldsymbol{x} \right) = \begin{bmatrix} K\left( \boldsymbol{x}, \textbf{rv}_1 \right) & ... & K\left( \boldsymbol{x}, \textbf{rv}_{N_{\text{rv}}}\right) \end{bmatrix}^T$ otherwise, $\textbf{rv}_j,j=1,...,N_{\text{rv}}$ are the relevance vectors found, $N_{\text{rv}}$ is the total number of relevance vectors, and the pruned $\boldsymbol{\Sigma}$ and $\boldsymbol{\mu}$ only contain components associated with relevance vectors. Then, the point estimate is taken as the posterior mean in Equation (\ref{eqn:mean_estimate}), and the three-sigma credible interval is obtained from the variance in Equation (\ref{eqn:var_estimate}), i.e., $\left(t-3\sigma,t+3\sigma\right)$.

Two practical customizations are adopted in this study when implementing the RVR model \cite{Tipping02}. First, the default settings from the sklearn-RVM package \cite{Tipping03}, including the initial values of $\alpha_i$ and $\beta^{-1}$, the pruning criterion, and the stopping conditions, are utilized. Second, to keep $\boldsymbol{\Sigma}$ and $\boldsymbol{\mu}$ well-conditioned throughout the iterations and to accelerate the convergence of Optimization (\ref{eqn:maximum_likelihood_opt}), all training data $\left\{ \left( \boldsymbol{x}_i, y_i \right) \right\}_{i=1}^N$ are standardized using Equation (\ref{eqn:standardize}) first before the RVR is trained. 
\section{BATTERY MODULE DATASET}
\label{Dataset}
This section describes the experimental module dataset used by this paper to develop and validate the proposed framework in Fig. \ref{fig:Algorithm}. The dataset is collected with modules consisting of Sony 18650 VTC6 cells that employ a lithium nickel-cobalt-aluminum oxide (NCA) positive electrode and a graphite-silicon composite negative electrode \cite{Lain}. From now on, these cells are referred to as NCA cells for brevity. The dataset is now publicly available, with a detailed description given in \cite{ZhouDatasetDrscriptor} and a download link provided in \cite{ZhouDataset}.

\subsection{Dataset Overview}
\label{DataOverview}
\begin{figure}[H]
    \centering
    \includegraphics[width=0.35\textwidth,trim=15 1 30 20,clip]{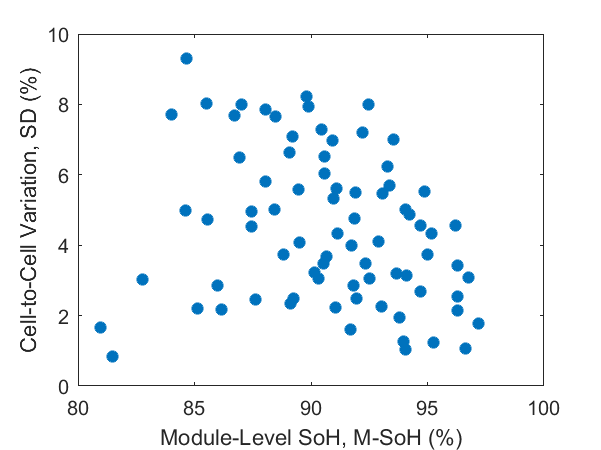}
    \caption{Distribution of Cell-to-Cell Variations and Module-Level SoHs inside Dataset}
    \label{fig:Data_Distribution}
\end{figure}

Each module consists of three NCA cells connected in parallel. Within each module, all NCA cells possess different C-SoH values, thereby introducing CtCVs. As a result, each module exhibits a distinct M-SoH and a unique degree of CtCV. The key attributes of the dataset are summarized in Table \ref{table:Datasets}. The distribution of the M-SoH and CtCV is shown in Fig. \ref{fig:Data_Distribution}. 

\begin{table}[H]
\centering
\caption{Key Attributes of the Battery Dataset}
\label{table:Datasets}
\begin{tabular}{|l|l|} 
\hline
\textbf{Attributes} & \textbf{Descriptions} \\ \hline\hline
Chemistry & NCA \cite{Lain} \\ \hline
Module Nominal Capacity & 9Ah \\ \hline
Module Configuration & 3 Cells in Parallel \\ \hline
No. of Modules & 78 \\ \hline
M-SoH Span & 100\% - 80.98\% \\ \hline
CtCV Span & 0\% - 9.31\% SD \\ \hline
C-Rates & 0.5C, 0.25C \\ \hline
No. of Total Datapoints & 156 \\ \hline
\end{tabular}
\end{table} 

Table \ref{table:IntuitiveComparisonOfMetrics} summarizes CtCV, M-SoH, and C-SoH values for several representative modules in the dataset. Based on Table \ref{table:IntuitiveComparisonOfMetrics}, the module with the maximum CtCV in the dataset exhibits a difference of approximately 21\% in the C-SoH values, while the module with the median CtCV still shows a difference of about 10\% in C-SoH values. Thus, the dataset encompasses modules with a broad range of CtCV levels. Table \ref{table:IntuitiveComparisonOfMetrics} also helps build intuition regarding the scale of SD as a CtCV metric.

\begin{table}[H]
\centering
\caption{Example Modules with Corresponding Module-Level SoH and Cell-to-Cell Variations}
\label{table:IntuitiveComparisonOfMetrics}
\begin{tabular}{|c|c|c|} 
\hline
\begin{tabular}[c]{@{}c@{}} \textbf{Cell-to-Cell} \\ \textbf{Variation} \end{tabular} & \begin{tabular}[c]{@{}c@{}} \textbf{Module-Level} \\ \textbf{SoH} \end{tabular} & \begin{tabular}[c]{@{}c@{}} \textbf{Cell-Level} \\ \textbf{SoH} \end{tabular} \\ \hline\hline
\begin{tabular}[c]{@{}c@{}} 9.31\% SD \\ (Maximum CtCV in Dataset) \end{tabular} & 84.65\% SoH & \begin{tabular}[c]{@{}c@{}} 99.2\% SoH \\ 80.6\% SoH \\ 78.5\% SoH  \end{tabular} \\ \hline
\begin{tabular}[c]{@{}c@{}} 4.54\% SD \\ (Median CtCV in Dataset) \end{tabular} & 88.67\% SoH & \begin{tabular}[c]{@{}c@{}} 91.9\% SoH \\ 91.8\% SoH \\ 82.3\% SoH \end{tabular} \\ \hline
\begin{tabular}[c]{@{}c@{}} 0.84\% SD \\ (Minimum CtCV in Dataset) \end{tabular} & 81.46\% SoH & \begin{tabular}[c]{@{}c@{}} 83.0\% SoH \\ 81.9\% SoH \\ 80.9\% SoH \end{tabular} \\ \hline
\end{tabular}
\end{table}

In summary, this dataset contains 78 distinct modules with M-SoH from 100\% to 80.98\% and CtCV from 0\% to 9.31\% SD. Therefore, the dataset covers a diverse range of scenarios and is well-suited for algorithm development, particularly for automotive applications where M-SoH typically spans from 100\% to 80\%.

\subsection{Experiment Protocols}
A detailed description of the experimental setup and protocols is provided in \cite{ZhouDatasetDrscriptor}, so this subsection only provides a brief overview. All characterization cycles for cells and modules consist of constant-current-constant-voltage (CC-CV) charging and constant-current (CC) discharging. 

First, 70 fresh cells are aged to form an inventory with diverse C-SoH levels ranging from 100\% to 79\%. Each aging cycle consists of 1.33C charging and 3.33C discharging. After aging, all the cells are characterized at 1C to obtain C-SoH labels and enable direct quantification of CtCV within modules. Then, 78 modules are assembled using off-the-shelf module holders, each comprising three parallel-connected cells with different C-SoH values. Then, the modules are characterized at 0.5C and 0.25C to obtain M-SoH labels, yielding 156 datapoints in total. These module characterization cycles will also be used for ICA/DVA later. As modules undergo repeated testing, cells within modules naturally age. To preserve accurate C-SoH and CtCV information, an additional re-characterization cycle at 1C is performed on cells according to a schedule. More details regarding the schedule of cell re-characterization are given in \cite{ZhouDatasetDrscriptor}.

\begin{remark}
    (On Robustness against Unmeasured Uncertainty) It should be noted that the interconnect resistances among different cells within each module are neither used nor characterized. Since such resistances are typically not measurable in onboard settings for real-world applications, a robust estimation algorithm should be able to provide accurate M-SoH and CtCV estimates without explicit knowledge of them. Accordingly, the proposed framework does not use interconnect resistances as inputs, and the absence of these measurements provides a realistic and meaningful validation of the algorithm under practical operating conditions. Nevertheless, these interconnect resistances may introduce a source of uncertainty for module-level performance characterization.
\end{remark}

\begin{figure}[H]
    \centering
    \begin{subfigure}[h]{0.48\textwidth}
        \centering
        \includegraphics[width=\textwidth,trim=10 80 69 15,clip]{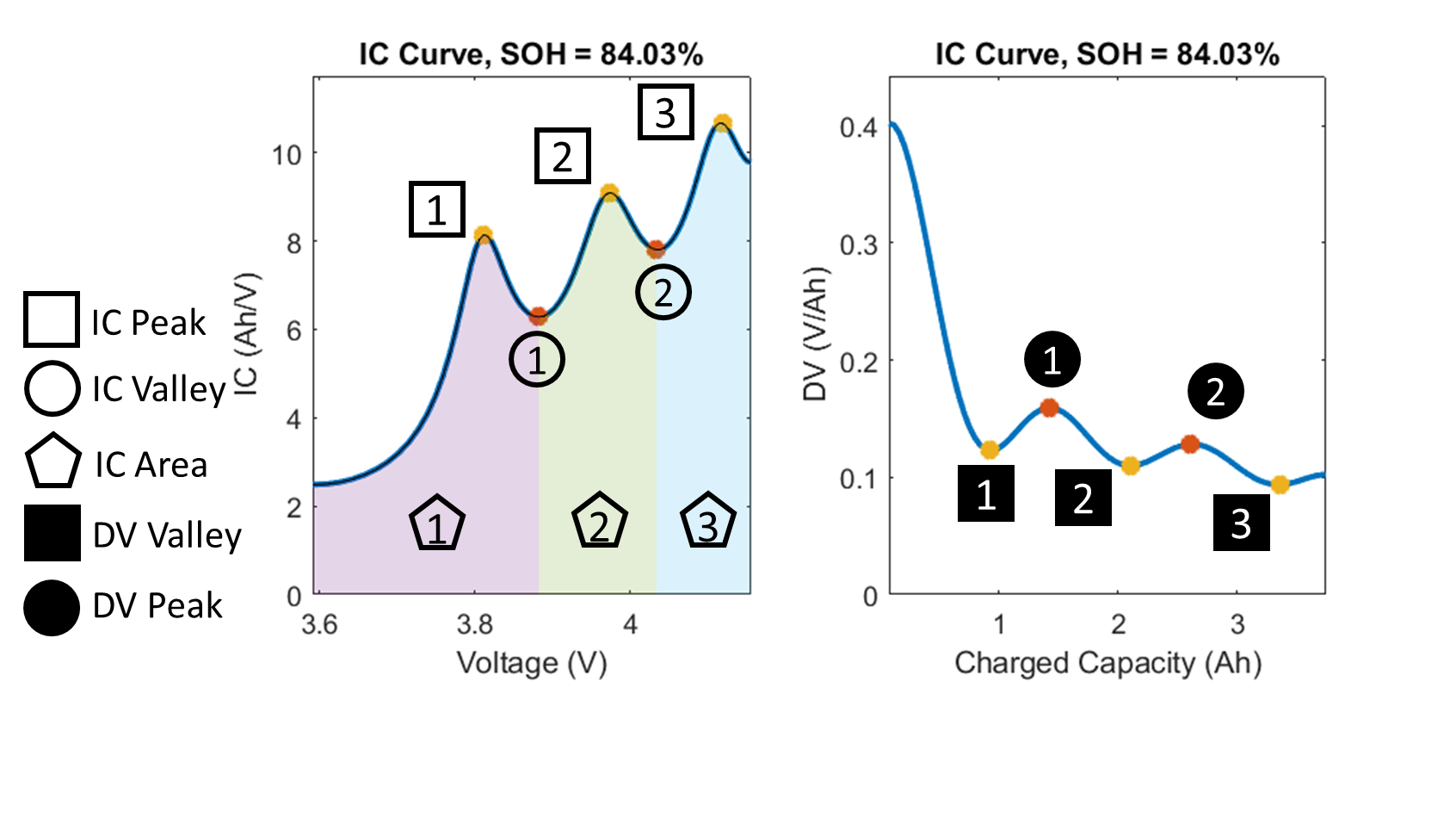}
        \caption{C-Rate = 0.5C}
        \label{fig:05C_Example_ICDVCurvesss}
    \end{subfigure} \hfill
    \begin{subfigure}[h]{0.48\textwidth}
        \centering
        \includegraphics[width=\textwidth,trim=10 80 69 15,clip]{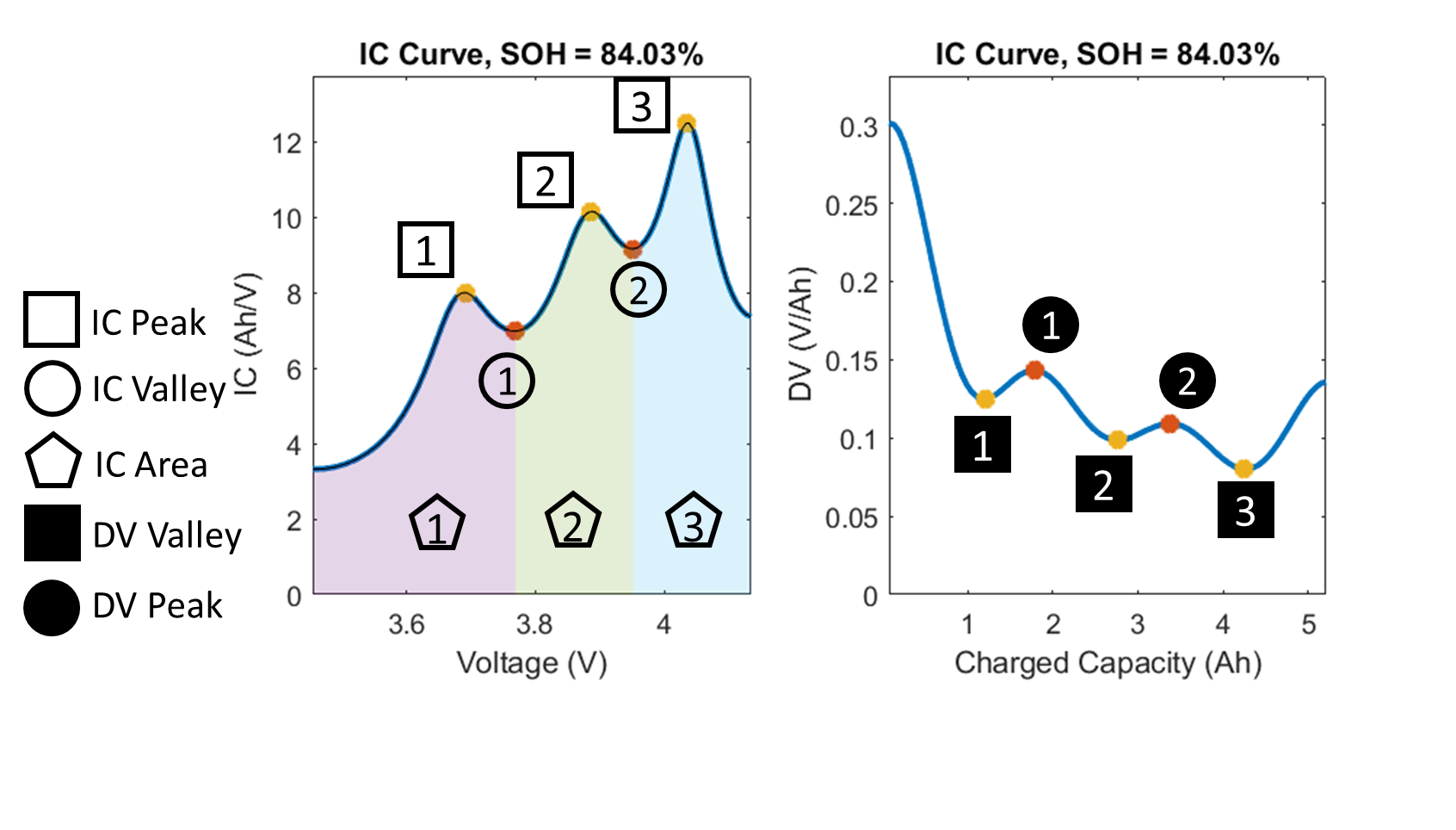}
        \caption{C-Rate = 0.25C}
        \label{fig:025C_Example_ICDVCurvesss}
    \end{subfigure} \hfill
\caption{Example M-IC/DV Curves and Related Features for the Dataset}
\label{fig:Dataset_Features}
\end{figure}

\subsection{Module-Level IC/DV Curves and Features}

\begin{figure*}[ht]
    \centering
    \begin{subfigure}[h]{0.75\textwidth}
        \centering
        \includegraphics[width=\textwidth,trim=5 5 5 5,clip]{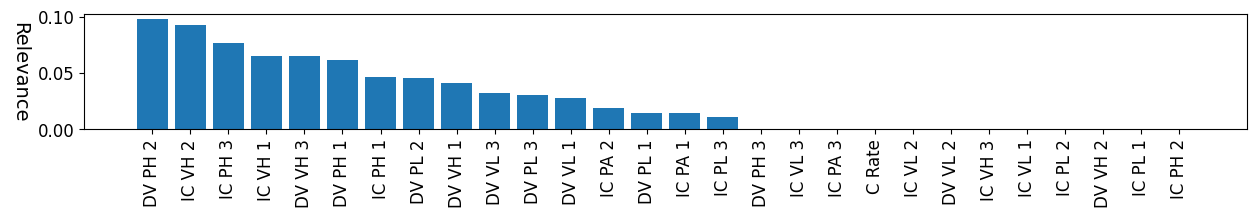}
        \caption{CtCV Estimation}
        \label{fig:CtCV_Relevance}
    \end{subfigure} \hfill

    \begin{subfigure}[h]{0.75\textwidth}
        \centering
        \includegraphics[width=\textwidth,trim=5 5 5 5,clip]{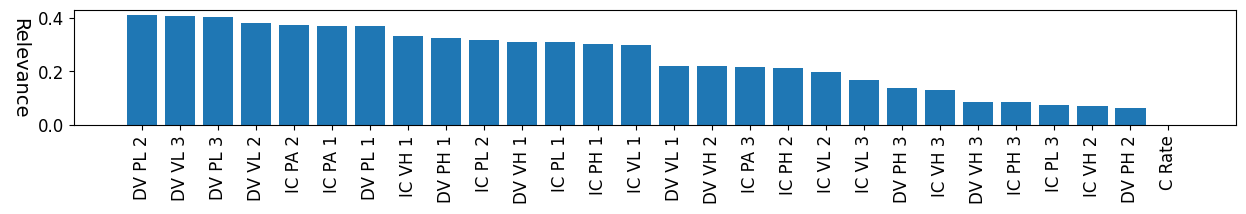}
        \caption{M-SoH Estimation}
        \label{fig:SOH_Relevance}
    \end{subfigure} \hfill
\caption{Feature Relevance for Cell-to-Cell Variation and Module-Level SoH Estimations}
\label{fig:FS_Relevance}
\end{figure*}

\begin{figure*}[ht]
    \centering
    \begin{subfigure}[h]{0.4\textwidth}
        \centering
        \includegraphics[width=\textwidth,trim=5 5 4 5,clip]{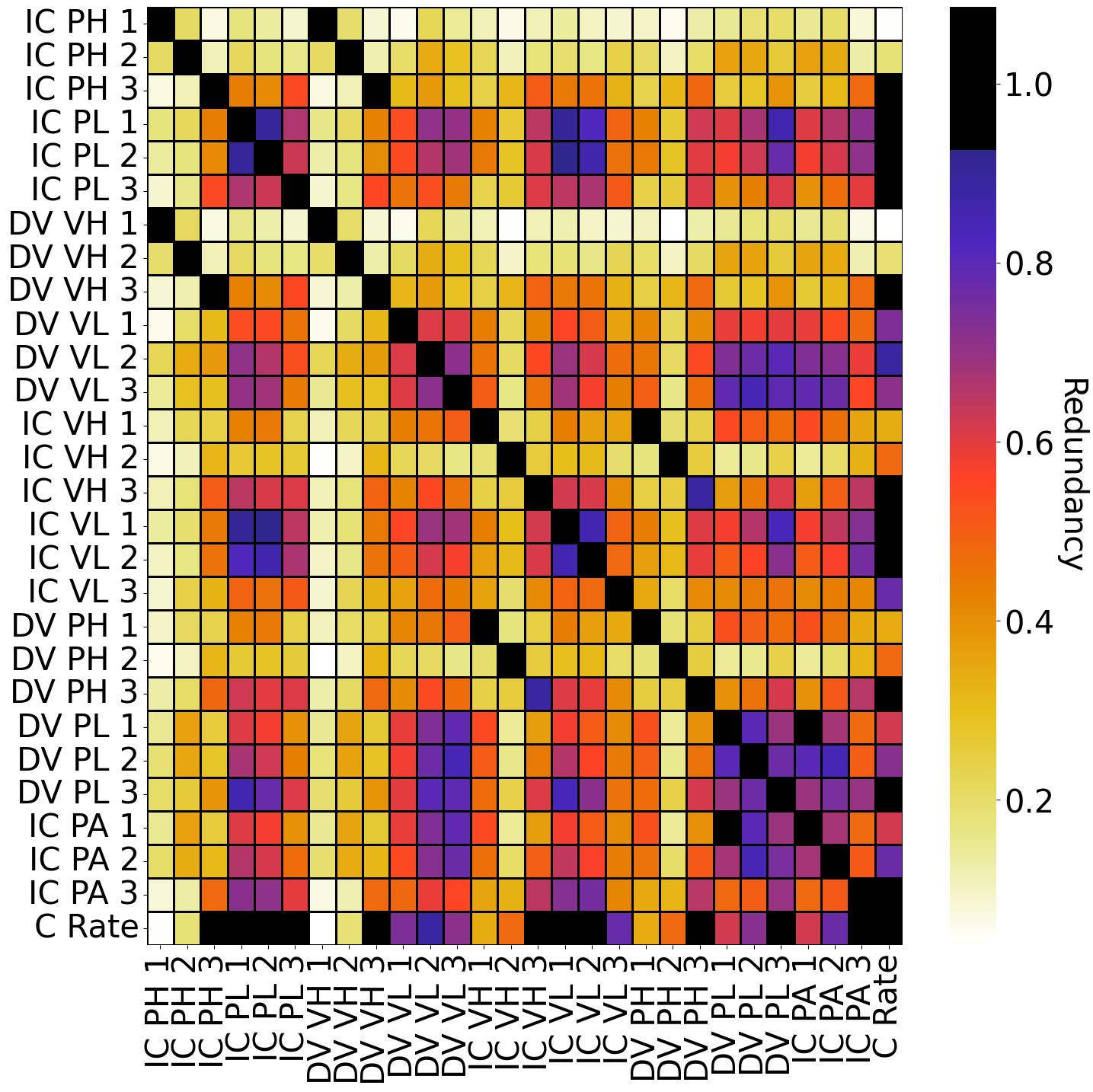}
        \caption{CtCV Estimation}
        \label{fig:CtCV_Redundancy}
    \end{subfigure}
    \begin{subfigure}[h]{0.4\textwidth}
        \centering
        \includegraphics[width=\textwidth,trim=5 5 4 5,clip]{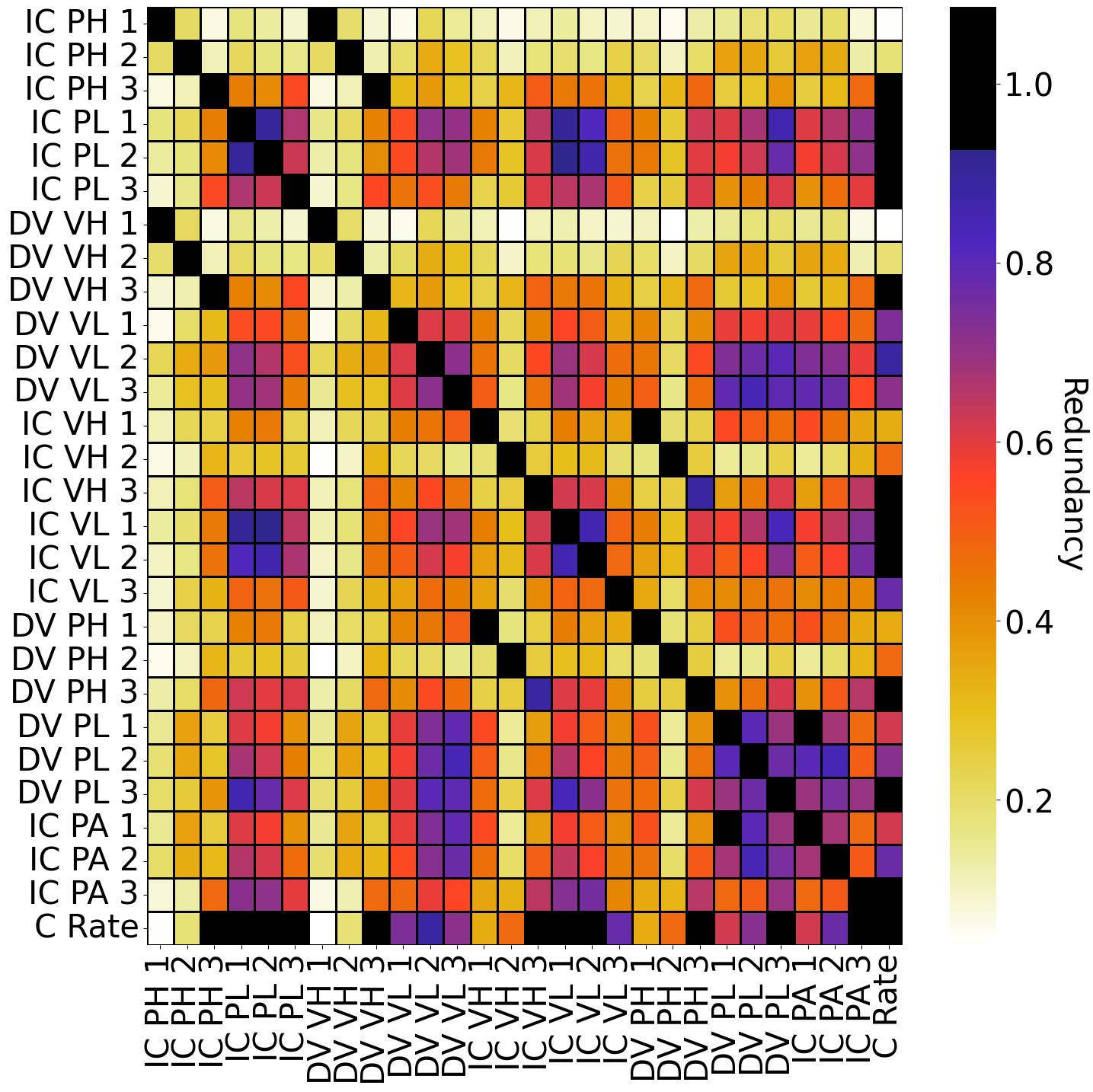}
        \caption{M-SoH Estimation}
        \label{fig:SOH_Redundancy}
    \end{subfigure}
\caption{Feature Redundancy for Cell-to-Cell Variation and Module-Level SoH Estimations}
\label{fig:FS_Redundancy}
\end{figure*}

\begin{figure*}[ht]
    \centering
    \begin{subfigure}[h]{0.4\textwidth}
        \centering
        \includegraphics[width=\textwidth,trim=5 5 4 5,clip]{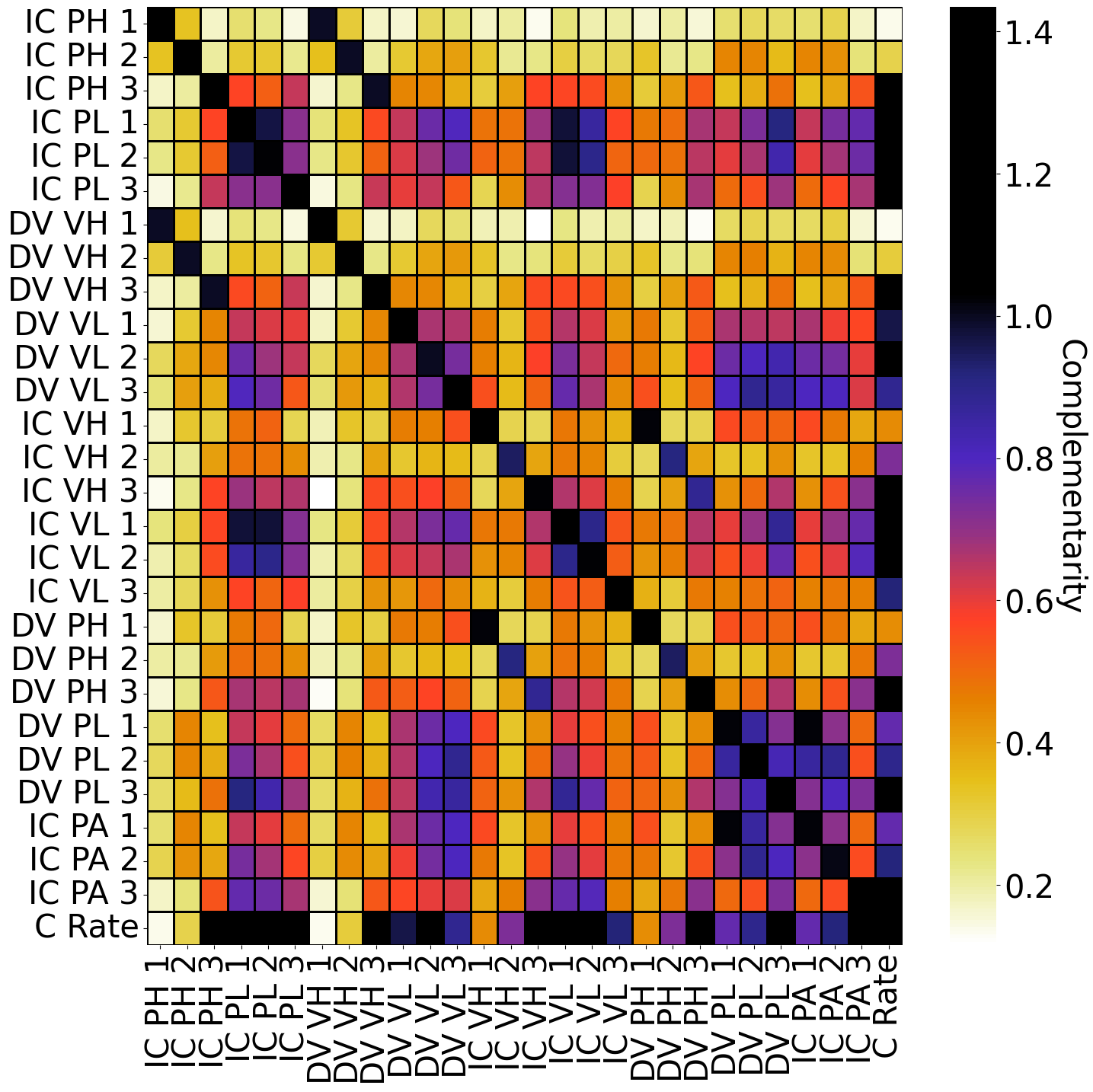}
        \caption{CtCV Estimation}
        \label{fig:CtCV_Complementarity}
    \end{subfigure}
    \begin{subfigure}[h]{0.4\textwidth}
        \centering
        \includegraphics[width=\textwidth,trim=5 5 4 5,clip]{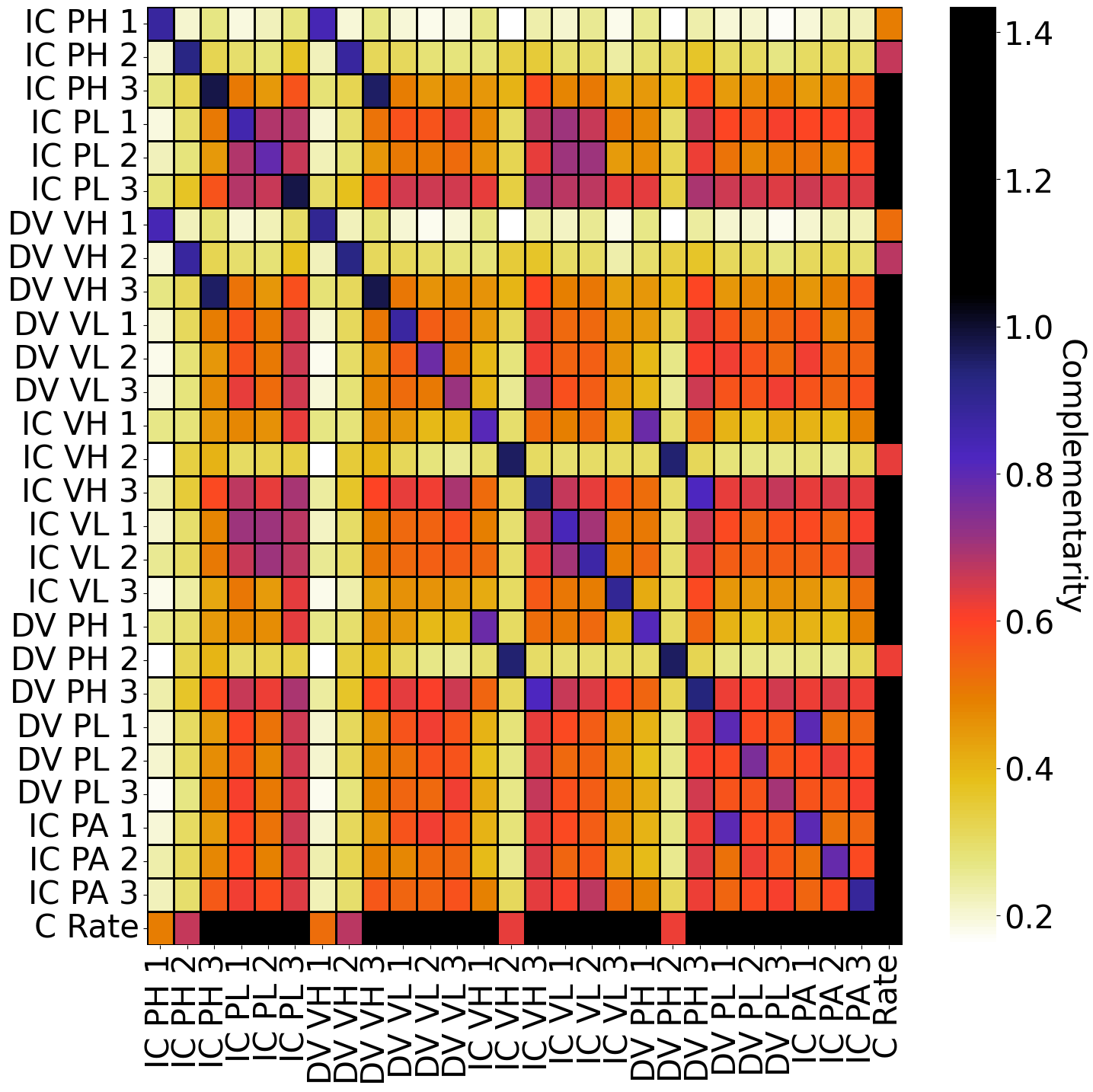}
        \caption{M-SoH Estimation}
        \label{fig:SOH_Complementarity}
    \end{subfigure}
\caption{Feature Complementarity for Cell-to-Cell Variation and Module-Level SoH Estimations}
\label{fig:FS_Complementarity}
\end{figure*}

\begin{table*}
\centering
\caption{Feature Ranking Results for Cell-to-Cell Variation Estimations}
\label{Table:FS_Results_CtCV}
\begin{tabular}{|l|llllllll|} 
\hline
\textbf{Ranked Selected Feature Set, $\mathcal{S}$} & \{ & DV PH 2, & C Rate, & DV PL 2, & DV VL 3, & DV VH 1, & DV VL 1, &  \\
 &  & IC VH 1, & DV VH 2, & DV PL 1, & IC PA 2, & IC VL 3, & DV VL 2 &  \} \\
\hline \hline
\textbf{Unranked Removed Feature Set, $\mathcal{R}$} & \{ & IC PH 1, & IC PH 2, & IC PH 3, & IC PL 1, & IC PL 2, & IC PL 3, &  \\
 &  & DV VH 3, & IC VH 2, & IC VH 3, & IC VL 1, & IC VL 2, & DV PH 1, &  \\
 &  & DV PH 3, & DV PL 3, & IC PA 1, &  IC PA 3 &  &  & \} \\
\hline
\end{tabular}
\end{table*}

\begin{table*}
\centering
\caption{Feature Ranking Results for Module-Level SoH Estimations}
\label{Table:FS_Results_SOH}
\begin{tabular}{|l|llllllll|} 
\hline
\textbf{Ranked Selected Feature Set, $\mathcal{S}$} & \{ & DV PL 2, & C Rate, & IC VH 1, & DV VH 1, & DV VL 3, & DV PL 1, &  \\
 &  & IC PA 2, & DV VL 2, & DV VH 2, & DV VL 1, & IC VL 3, & IC VH 2 & \}  \\
\hline \hline
\textbf{Unranked Removed Feature Set, $\mathcal{R}$} & \{ & IC PH 1, & IC PH 2, & IC PH 3, & IC PL 1, & IC PL 2, & IC PL 3, &  \\
 &  & DV VH 3, & IC VH 3, & IC VL 1, & IC VL 2, & DV PH 1, & DV PH 2, &  \\
 &  & DV PH 3, & DV PL 3, & IC PA 1, &  IC PA 3 &  &  & \} \\
\hline
\end{tabular}
\end{table*}

Fig. \ref{fig:Dataset_Features} shows example M-IC/DV curves and related features under 0.5C and 0.25C. Note that, for ease of discussion, the acronyms in Table \ref{table:Features} and indices labeled in Fig. \ref{fig:Dataset_Features} will be used together to refer to different features. For example, the left IC peak height in Fig. \ref{fig:05C_Example_ICDVCurvesss} is denoted as IC PH 1 in Section \ref{Results}.

\begin{remark}
    (On Feature Disappearance under Severe Degradation) Planned for automotive applications, this framework was developed and validated for SoH up to 80\%, which is commonly regarded as the end of service life. Accordingly, this dataset only includes battery modules spanning 100\% to 80\% SoH, within which all important IC/DV features remain observable. However, important IC/DV features may disappear under severe degradation, especially in second-life batteries with $\text{SoH} < 80\%$.
\end{remark}

\section{PERFORMANCE OF THE PROPOSED FRAMEWORK}
\label{Results}
This section evaluates the performance of the proposed framework in Fig. \ref{fig:Algorithm} for CtCV and M-SoH estimations. A nested cross-validation approach \cite{Berrar} is used to train, validate, and test the proposed framework, where the inner loop performs ten-fold cross-validation to fine-tune hyperparameters and the outer loop performs leave-one-out cross-validation to examine the final performance. Note that leave-one-out cross-validation ensures that the proposed framework is tested across all 156 distinct scenarios (78 modules under 2 types of C-rates). Because each module has a unique combination of M-SoH and CtCV (as demonstrated in Section \ref{DataOverview}), every module represents a distinct unseen scenario in each iteration of the leave-one-out cross-validation.

As outlined in Sections \ref{InfoDefn} and \ref{RVR}, CtCV, M-SoH, and all the features are first standardized based on the training set using Equation (\ref{eqn:standardize}). Then, during result discussions, all these standardized quantities are converted back for physical interpretations.

\subsection{Feature Ranking and Selection Results} 
\label{FS_Results}
Fig. \ref{fig:FS_Relevance} ranks the relevance of all the features for both CtCV and M-SoH estimations. Based on Fig. \ref{fig:FS_Relevance}, three key observations can be made: (i) the C-rate exhibits zero relevance to both CtCV and M-SoH, confirming the argument in Section \ref{FS_Algm}; (ii) DV PH 2 and DV PL 2 exhibit the highest relevance to CtCV and M-SoH, respectively, and are therefore selected first when no features are pre-selected; and (iii) all M-IC/DV features exhibit very low relevance to CtCV, consistent with the complex distortion patterns discussed in Section \ref{Distortion}. This indicates that CtCV estimation cannot rely on direct feature-output correlations and must instead primarily exploit interactions among multiple features, underscoring the importance of incorporating redundancy and complementarity criteria in feature selection.

Fig. \ref{fig:FS_Redundancy} illustrates the redundancy among all features, where black denotes redundancy values above the threshold $\tilde{I}_{\text{th}}$. As shown in Fig. \ref{fig:FS_Redundancy}, the proposed redundancy criterion (i) correctly identifies physically redundant features, such as IC peak/valley heights versus DV valley/peak heights and DV PL 1 versus IC AR 1, confirming its effectiveness, and (ii) reveals fully redundant feature pairs that are not readily inferred from physical insight, such as the C-rate being redundant with multiple IC/DV features. These data-driven redundancies are specific to the present dataset and may not generalize across different battery chemistries or operating conditions. Nonetheless, the proposed feature selection algorithm remains unchanged and fully generalizable to different datasets without modification.

Fig. \ref{fig:FS_Complementarity} shows the complementarity among all features. Based on Fig. \ref{fig:FS_Complementarity}, the C-rate has very strong complementarity with most IC/DV features for CtCV and M-SoH estimations, aligning with the physical knwoledge and the argument in Section \ref{FS_Algm}. Thus, in the case where these M-IC/DV features are selected, the C rate will be selected based on Equation (\ref{eqn:FS_obj}), unless the C-rate is completely redundant to the associated feature. Note that completely redundant features also have high CMI, but they will not be selected because of the feature removal process in Algorithm \ref{alg:infoFS}. In contrast, the complementarities among most other IC/DV features do not exhibit obvious and direct physical interpretations and are therefore dataset-dependent.

With the feature relevance, redundancy, and complementarity computed, Algorithm \ref{alg:infoFS} produces the sets of selected and removed features for both CtCV and M-SoH estimations, as summarized in Table \ref{Table:FS_Results_CtCV} and Table \ref{Table:FS_Results_SOH}, respectively. As shown in these tables, the optimal feature sets differ between CtCV and M-SoH estimations. Notably, however, the C-rate is ranked highly in both cases due to its strong complementarity, as discussed earlier.

\subsection{Performance of Cell-to-Cell Variation and Module-Level SoH Estimation} 
\label{Estimation_Results}

\begin{figure*}[ht]
\centering
\begin{subfigure}{0.3\textwidth}
\includegraphics[width=\textwidth,trim=4 5 4 5,clip]{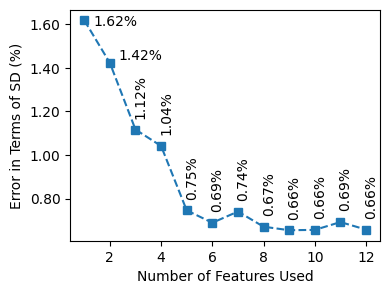}
\caption{Estimation MAE}
\label{fig:CtCV_MAE}
\end{subfigure}\hfill
\begin{subfigure}{.3\textwidth}
\includegraphics[width=\textwidth,trim=4 5 4 5,clip]{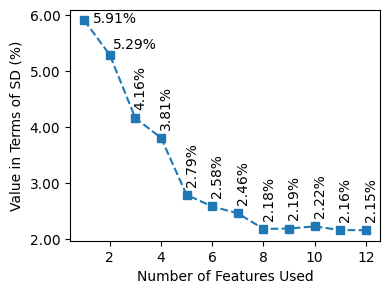}
\caption{Average Three-Sigma Values}
\label{fig:CtCV_3Sigma}
\end{subfigure}\hfill
\begin{subfigure}{.3\textwidth}
\includegraphics[width=\textwidth,trim=4 5 4 5,clip]{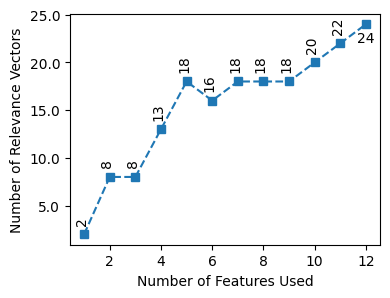}
\caption{Number of Relevance Vectors}
\label{fig:CtCV_N}
\end{subfigure}
\caption{Cell-to-Cell Variation Estimation Performance under Different Numbers of Features Used}
\label{fig:CtCV_Results}
\end{figure*}

\begin{figure*}[ht]
\centering
\begin{subfigure}{0.3\textwidth}
\includegraphics[width=\textwidth,trim=4 5 4 5,clip]{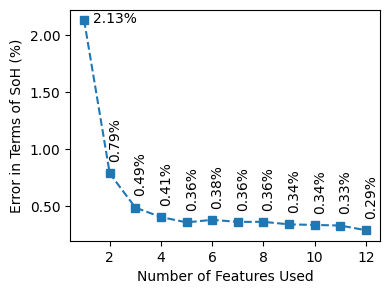}
\caption{Estimation MAE}
\label{fig:SOH_MAE}
\end{subfigure}\hfill
\begin{subfigure}{.3\textwidth}
\includegraphics[width=\textwidth,trim=4 5 4 5,clip]{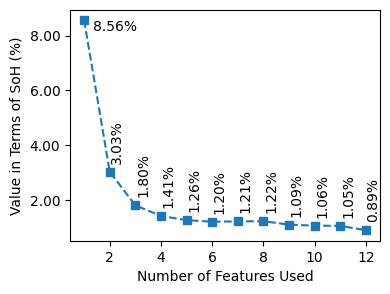}
\caption{Average Three-Sigma Values}
\label{fig:SOH_3Sigma}
\end{subfigure}\hfill
\begin{subfigure}{.3\textwidth}
\includegraphics[width=\textwidth,trim=4 5 4 5,clip]{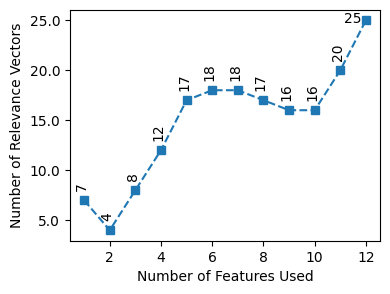}
\caption{Number of Relevance Vectors}
\label{fig:SOH_N}
\end{subfigure}
\caption{Module-Level SoH Estimation Performance under Different Numbers of Features Used}
\label{fig:SOH_Results}
\end{figure*}

To reflect the trade-off between estimation performance and model complexity, Fig. \ref{fig:CtCV_Results} and Fig. \ref{fig:SOH_Results} demonstrate the final testing mean absolute error (MAE), numbers of relevance vectors, and average three-sigma values when different numbers of features are used. The corresponding numerical values are also annotated in these figures. Both figures exhibit a consistent trend: for both CtCV and M-SoH estimations, both testing MAEs and estimation uncertainties (three-sigma values) decrease, as additional optimally selected features are incorporated, but eventually plateau once the most informative features have been included. After reaching this plateau, small fluctuations in MAEs and three-sigma values may occur due to statistical variability. The model complexity (i.e., the number of relevance vectors), on the other hand, generally increases as the number of features grows, although it is not strictly monotonic and may decrease for certain feature subsets. Thus, these plots provide a practical guideline for selecting an appropriate number of features depending on the desired balance between accuracy and model complexity. 

\begin{figure}[h]
\centering
\begin{subfigure}{0.3\textwidth}
\includegraphics[width=\textwidth,trim=4 5 4 5,clip]{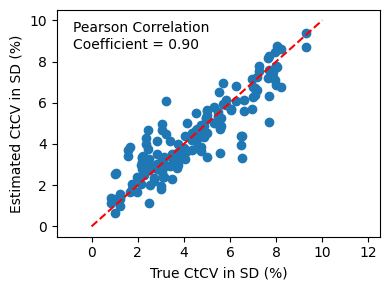}
\caption{CtCV Estimation}
\label{fig:CtCV_CaseStudy}
\end{subfigure}\hfill
\begin{subfigure}{0.3\textwidth}
\includegraphics[width=\textwidth,trim=4 5 4 5,clip]{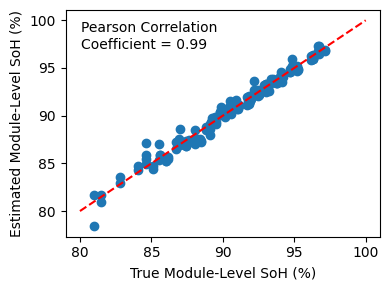}
\caption{M-SoH Estimation}
\label{fig:SOH_CaseStudy}
\end{subfigure}\hfill
\caption{Distribution between Estimation and Ground Truth}
\label{fig:CaseStudy}
\end{figure}

To provide specific performance examples, Fig. \ref{fig:CaseStudy} shows the distributions of estimation versus ground truth when 6 and 4 features are used for CtCV estimation and M-SoH estimation, respectively. Based on Fig. \ref{fig:CtCV_Results}, Fig. \ref{fig:SOH_Results}, and Fig. \ref{fig:CaseStudy}, several important conclusions could be made: 
\begin{itemize}
    \item Both CtCV and M-SoH can be estimated using only module-level information such as M-IC/DV features. Better performance can be achieved at the expense of higher model complexity.  
    \item CtCV and M-SoH can be estimated independently from each other. Note that the proposed method does not use any information about CtCV explicitly for M-SoH estimation, nor any information about M-SoH for CtCV estimation. The proposed framework only uses module-level IC/DV features and charging conditions.
\end{itemize}

\begin{figure}[h]
\centering
\begin{subfigure}{0.3\textwidth}
\includegraphics[width=\textwidth,trim=4 5 4 5,clip]{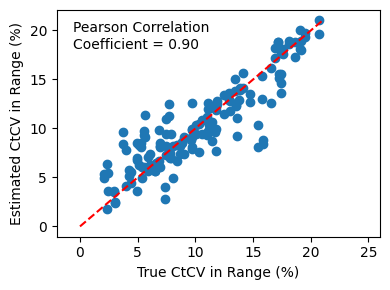}
\caption{Range Estimation}
\label{fig:CtCV_CaseStudy_Range}
\end{subfigure}\hfill
\begin{subfigure}{0.3\textwidth}
\includegraphics[width=\textwidth,trim=4 5 4 5,clip]{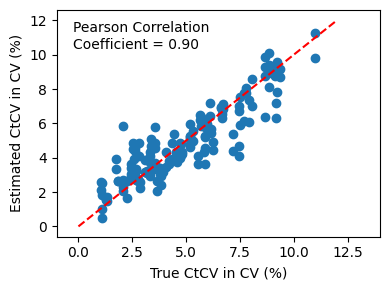}
\caption{Coefficient of Variation Estimation}
\label{fig:CtCV_CaseStudy_CV}
\end{subfigure}\hfill
\caption{Performance of Proposed Framework on Estimating Other Cell-to-Cell Variation Metrics}
\label{fig:CaseStudy_OtherMetrics}
\end{figure}

\begin{remark}
    (On Applying the Proposed Framework to Other CtCV Metrics) As discussed in Section \ref{Metric}, no universally accepted metric exists for quantifying CtCVs. The proposed framework is flexible and can estimate the CtCV quantified by other metrics besides population SD. For illustration, consider the other two commonly used metrics, namely the range and the coefficient of variation (CV) \cite{Wildfeuer, Fernández}, defined as: 
    \begin{eqnarray}
        \text{Range} &=& \text{max} \left( \left\{ \text{C-SoH}_i \right\} \right) - \text{min} \left( \left\{ \text{C-SoH}_i \right\} \right) \\ 
        \text{CV} &=& \frac{\text{sd} \left( \left\{ \text{C-SoH}_i \right\} \right)}{\text{mean} \left( \left\{ \text{C-SoH}_i \right\} \right)} = \frac{\text{SD}}{\text{M-SoH}} 
    \end{eqnarray}
    Different CtCV metrics will lead to different feature ranking, resulting in different optimal feature sets. Fig. \ref{fig:CaseStudy_OtherMetrics} shows the estimation performance for range (Fig. \ref{fig:CtCV_CaseStudy_Range}) and CV (Fig. \ref{fig:CtCV_CaseStudy_CV}) when the top 6 features are used. Comparing the Pearson correlation coefficients between true and estimated CtCVs in Fig. \ref{fig:CaseStudy} and \ref{fig:CaseStudy_OtherMetrics} demonstrates that the proposed framework achieves consistently strong performance, confirming its good generalizability to various CtCV metrics.
\end{remark}

\subsection{Onboard Implementation of the Proposed Framework} 
The proposed method in Fig. \ref{fig:Algorithm} has both offboard and onboard computations involved. The proposed feature selection algorithm in Step 3 of Fig. \ref{fig:Algorithm} is performed completely offboard to find the optimal set of features for both CtCV and M-SoH estimations. The onboard computation involves: (i) extracting the values of these selected features from M-IC/DV curves and (ii) estimating CtCV and M-SoH according to the RVR models developed in Step 4 of Fig. \ref{fig:Algorithm}. 

The proposed RVR-based estimation models are sparse, namely, a small number of relevance vectors and one offset scalar need to be stored onboard. Only Equations (\ref{eqn:mean_estimate}) and (\ref{eqn:var_estimate}) need to be computed onboard for each module. These two equations involve low-dimensional matrix multiplications and do not require any computationally intensive operations (e.g., matrix inversion). Take the NCA modules as an example. If one uses 4 features to estimate M-SoH, according to Fig. \ref{fig:SOH_Results}, twelve 4-by-1 relevance vectors and one offset scalar will be stored onboard and used across all modules. Then, for each module, Equations (\ref{eqn:mean_estimate}) and (\ref{eqn:var_estimate}) involve matrix multiplications among four 13-by-1 vectors and one 13-by-13 matrix. Similarly, if one uses 6 features to estimate CtCV, according to Fig \ref{fig:CtCV_Results}, sixteen 6-by-1 relevance vectors and one offset scalar will be stored onboard and used across all modules. Then, for each module, Equations (\ref{eqn:mean_estimate}) and (\ref{eqn:var_estimate}) involve matrix multiplications among four 17-by-1 vectors and one 17-by-17 matrix.

Given its low computational complexity, the proposed framework can be readily integrated into battery management systems and implemented on corresponding hardware. From a system integration perspective, the proposed framework requires only standard module-level measurements that are already available in practice battery management systems, namely module-level voltage, current, and temperature during charging. Therefore, no additional sensing hardware is required.

In summary, the onboard computational footprint of the proposed method in Fig. \ref{fig:Algorithm} is small and involves only extracting feature values and performing low-dimensional matrix multiplications. All the computationally intensive optimization and training processes are done offboard. Moreover, module-level degradation monitoring does not need to be performed continuously. Thus, its implementation will not impose special onboard computational requirements. 

\section{CONCLUSIONS}
\label{Conclusion}
This paper proposes a unified ICA/DVA-based framework for the estimation of CtCV and M-SoH for battery modules with parallel-connected cells, using only module-level measurements. By integrating information theory-based feature selection with relevance vector regression, the framework identifies the most informative module-level features and provides sparse estimation models. The performance of the proposed framework is evaluated using an experimental dataset that consists of two charging C-rates (0.25C and 0.5C) and 78 NCA modules spanning M-SoH from 100\% to 80.98\% and CtCV from 0\% to 9.31\% SD. Evaluation results demonstrate the following important conclusions.

First, the proposed framework provides accurate CtCV and M-SoH estimates with low computational complexity across different charging C-rates. Second, with the proposed framework, CtCV and M-SoH estimations can be performed independently. The proposed framework requires no CtCV information for M-SoH estimation and no M-SoH information for CtCV estimation. This allows each problem to be addressed using dedicated algorithms without mutual interference and provides more freedom in estimation algorithm design. Third, the framework is tested for different CtCV metrics. Comparable accuracy is obtained for different metrics, such as SD, CV, and range. Overall, this work establishes the first experimentally validated and computationally efficient framework capable of quantitatively estimating both CtCV and M-SoH for modules under varying C-rates using only module-level measurements.

The future work involves improving generalizability, robustness, and scalability of the proposed framework. First, regarding generalizability, one remaining issue is feature availability. Some features may disappear under severe degradation, while others may be unavailable under narrower or different charging voltage windows. However, the proposed framework does not yet provide a systematic way to handle such feature disappearance or unavailability. Another open question regarding generalizability is whether the framework can be extended to estimate CtCV in other cell-level properties, such as internal resistance. Second, for robustness, the effects of measurement noise, data acquisition systems, and vehicle chronometrics on the proposed framework need to be systematically assessed. Third, for scalability, the framework has only been validated on modules with three parallel-connected cells, but modules may contain more parallel-connected cells. It remains unclear whether accurate estimation can still be maintained as the number of parallel-connected cells increases.






\bibliographystyle{ieeeconf.bst}
\bibliography{ieeeconf.bib}
\end{document}